\documentclass[sigconf]{acmart}
\usepackage{booktabs,multirow}
\usepackage{enumitem}
\usepackage{kotex}
\usepackage{color}
\usepackage{rotating}

\newcommand{\ie}{{\it i.e.}}
\newcommand{\eg}{{\it e.g.}}
\newcommand{\com}{\textcolor{red}}
\newcommand{\fcom}{\textcolor{blue}}

\newcommand{\argmin}{\operatornamewithlimits{argmin}}

\newcommand{\ul}{\underline}{}

\copyrightyear{2021}
\acmYear{2021}
\setcopyright{iw3c2w3}
\acmConference[WWW '21]{Proceedings of the Web Conference 2021}{April 19--23, 2021}{Ljubljana, Slovenia}
\acmBooktitle{Proceedings of the Web Conference 2021 (WWW '21), April 19--23, 2021, Ljubljana, Slovenia}
\acmPrice{}
\acmDOI{10.1145/3442381.3450005}
\acmISBN{978-1-4503-8312-7/21/04}
\begin{document}
\title{Session-aware Linear Item-Item Models for Session-based Recommendation}

\author{Minjin Choi}
\affiliation{
  \institution{Sungkyunkwan University\\ Republic of Korea}}
\email{zxcvxd@skku.edu}

\author{Jinhong Kim}
\affiliation{
  \institution{Sungkyunkwan University\\ Republic of Korea}}
\email{legend7811@skku.edu}

\author{Joonseok Lee}
\affiliation{
  \institution{Google Research, USA\\ Seoul National University\\ Republic of Korea}}
\email{joonseok2010@gmail.com}

\author{Hyunjung Shim}
\affiliation{
  \institution{Yonsei University\\ Republic of Korea}}
\email{kateshim@yonsei.ac.kr}

\author{Jongwuk Lee}\authornote{Corresponding author}
\affiliation{
  \institution{Sungkyunkwan University\\ Republic of Korea}}
\email{jongwuklee@skku.edu}

\begin{abstract}
Session-based recommendation aims at predicting the next item given a sequence of previous items consumed in the session, \eg, on e-commerce or multimedia streaming services. Specifically, session data exhibits some unique characteristics, \ie, \emph{session consistency} and \emph{sequential dependency} over items within the session, \emph{repeated item consumption}, and \emph{session timeliness}. In this paper, we propose simple-yet-effective linear models for considering the holistic aspects of the sessions. The comprehensive nature of our models helps improve the quality of session-based recommendation. More importantly, it provides a generalized framework for reflecting different perspectives of session data. Furthermore, since our models can be solved by closed-form solutions, they are highly scalable. Experimental results demonstrate that the proposed linear models show competitive or state-of-the-art performance in various metrics on several real-world datasets.
\end{abstract}

\ccsdesc[500]{Information systems~Recommender systems}
\ccsdesc[500]{Information systems~Collaborative filtering}
\ccsdesc[300]{Information systems~Expert systems}

\keywords{Collaborative filtering; Session-based recommendation; Item similarity; Item transition; Closed-form solution}

\maketitle

\section{Introduction}

Recommender systems have been widely used to help users mitigate information overload and render useful information in various applications, such as e-commerce (\eg, Amazon and Alibaba) and on-line streaming services (\eg, YouTube, Netflix, and Spotify). Conventional recommender systems~\cite{GoldbergNOT92,HerlockerKBR99,HuKV08,RicciRS15,SedhainMSX15,HeLZNHC17,LeeBKLS14,LeeKLSB16,LeeAVN18,ChoiJLL21} usually personalize on user accounts, assumed to be owned by a single person and to be static over time. However, this assumption is often invalid. First of all, basic user information, \eg, demographic data, may not be verified. The same account may also be shared by multiple individuals, \eg, mixing browsing and click behaviors across family members. Even the same user may use her account in different ways depending on the context, \eg, work-related \emph{v.s} entertainment purpose. Therefore, relying purely on user accounts may result in sub-optimal personalization.

\begin{figure}
  \includegraphics[width=0.45\textwidth]{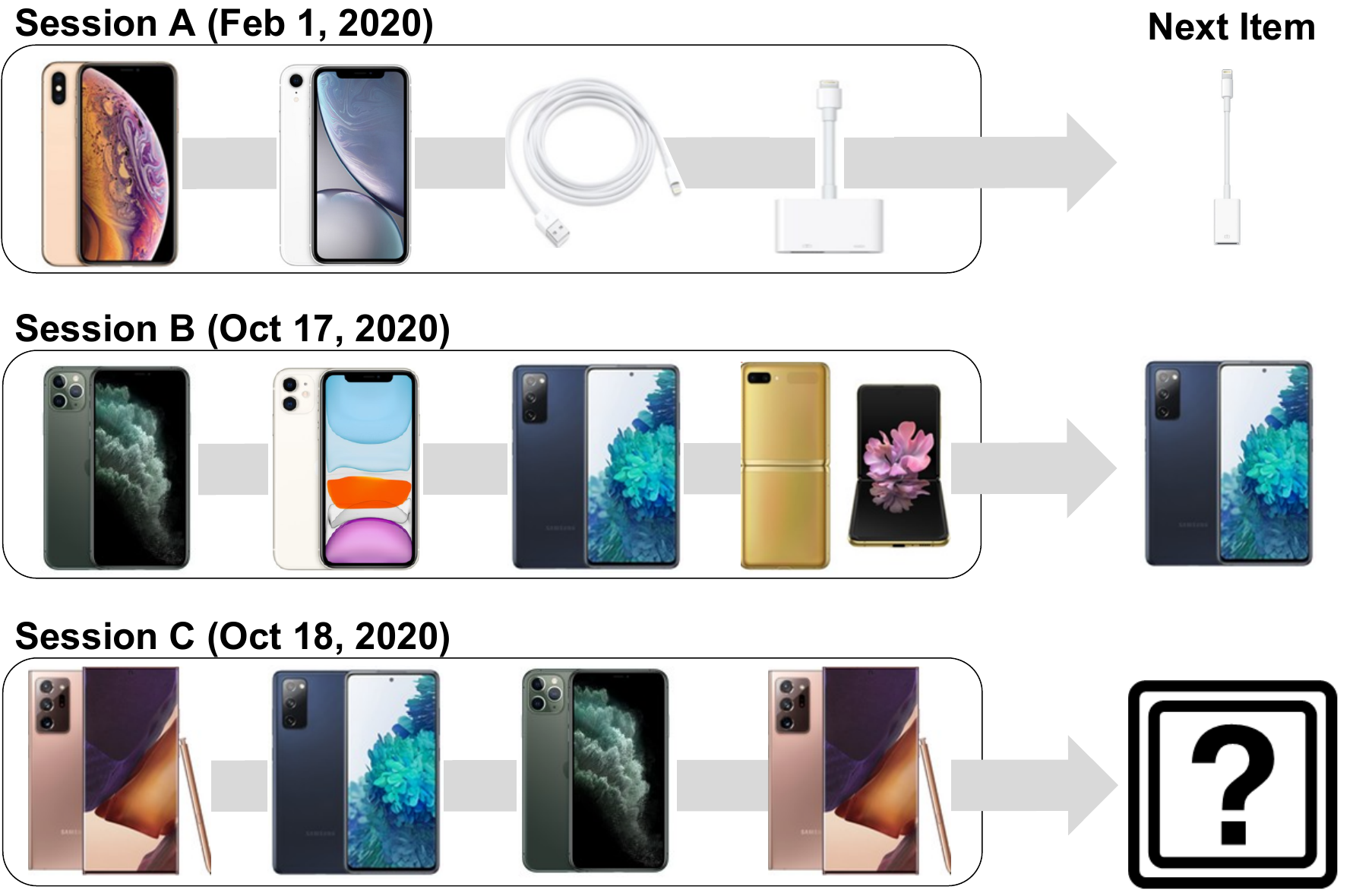}
  \caption{Illustration of various session properties, \ie, session consistency, sequential dependency, repeated item consumption, and timeliness of sessions.
  }
  \label{fig:figure1}
  \vskip -0.18in
\end{figure}

Recently, \emph{session-based recommendation}~\cite{BonninJ14, JannachLL17, WangCW19, FangGZS19, QuadranaCJ18}, where the recommendation relies solely on the user's actions in an ongoing session, has gained much attention to overcome the issues above. Compared to the traditional ones, session-based recommender systems exhibit some unique characteristics. Firstly, the items observed in the same session are often highly coherent and consistent with the user's specific intent, \eg, a list of products in the same category or a list of songs with a similar mood, referred to as \emph{session consistency}. The brand-new smartphones in ``session B'' in Figure~\ref{fig:figure1}, for example, are highly correlated. Secondly, some items tend to be consumed in a specific order, namely \emph{sequential dependency}, \eg, consecutive episodes of a TV series. As in ``session A'' of Figure~\ref{fig:figure1}, smartphone accessories are usually followed by smartphones, but not in the other way around. Thirdly, the user might repeatedly consume the same items in a session, called \emph{repeated item consumption}. For instance, the user may listen to her favorite songs repeatedly or choose the same smartphone for the comparisons as illustrated in ``session C'' of Figure~\ref{fig:figure1}. Lastly, recent sessions are generally a stronger indicator of the user's interest, namely, \emph{timeliness of sessions}. In Figure ~\ref{fig:figure1}, ``Session B'' and ``session C'' are close in time and share several popular items. The above four properties do not necessarily appear in all sessions, and one property may be dominant than others.

Recent session-based recommender systems have shown outstanding performance by utilizing deep neural networks (DNNs). Recurrent neural networks~\cite{HidasiKBT15,HidasiQKT16,HidasiK18} or attention mechanism~\cite{LiRCRLM17,LiuZMZ18} are applied to model sequential dependency, and graph neural networks (GNNs)~\cite{WuT0WXT19,XuZLSXZFZ19,GuptaGMVS19,PanCLR20a} are effective for representing session consistency. However, they mostly focus on some aspects of sessions and thus do not generalize well to various datasets. No single model can guarantee competitive performances across various datasets, as reported in Section~\ref{sec:results}. Furthermore, they generally require high computational cost for model training and inference.

To overcome the scalability issue of DNN-based models, recent studies~\cite{LudewigJ18, GargGMVS19} suggest neighborhood-based models for session-based recommendation, which are highly scalable. Surprisingly, they also achieve competitive performance comparable to DNN-based models on several benchmark datasets~\cite{LudewigMLJ19a, LudewigMLJ19b}. However, the neighborhood-based models only exploit neighboring sessions, limited to capture global patterns of sessions.

In this paper, we propose novel \emph{session-aware linear models}, to complement the drawback of DNN-based and neighborhood-based models. Specifically, we design a simple-yet-effective model that (i) comprehensively considers various aspects of session-based recommendations (ii) and simultaneously achieves scalability. The idea of the linear models has been successfully applied to traditional recommender systems (\eg, SLIM~\cite{NingK11} and EASE$^{R}$~\cite{Steck19b}). Notably, EASE$^{R}$~\cite{Steck19b} has shown impressive performance gains and enjoyed high scalability thanks to its closed-form solution.

Inspired by the recent successful studies~\cite{Steck19a,Steck19b,Steck19c,JeunenBG20}, we reformulate the linear model that captures various characteristics of sessions. Firstly, we devise two linear models focusing on different properties of sessions: (i) \emph{Session-aware Linear Item Similarity (SLIS)} model aims at better handling session consistency, and (ii) \emph{Session-aware Linear Item Transition (SLIT)} model focuses more on sequential dependency. With both SLIS and SLIT, we relax the constraint to incorporate repeated items and introduce a weighting scheme to take the timeliness of sessions into account. Combining these two types of models, we then suggest a unified model, namely \emph{Session-aware Item Similarity/Transition (SLIST)} model, which is a generalized solution to holistically cover various properties of sessions. Notably, SLIST shows competitive or state-of-the-art performance consistently on various datasets with different properties, proving its generalization ability. Besides, both SLIS and SLIT (and thus the combined SLIST) are solved by closed-form equations, leading to high scalability.


To summarize, the key advantages of SLIST are presented as follows: (i) It is a generalized solution for session-based recommendation by capturing various properties of sessions. (ii) It is highly scalable thanks to closed-form solutions of linear models. (iii) Despite its simplicity, it achieves competitive or state-of-the-art performance in various metrics (\ie, HR, MRR, Recall, and MAP) on several benchmark datasets (\ie, YouChoose, Diginetica, RetailRocket, and NowPlaying).

\section{Preliminaries}\label{sec:preliminaries}

We start with defining notations used in this paper, followed by a formal problem statement and review of linear item-item models.

\vspace{1mm}
\noindent
\textbf{Notations.} Let $\mathcal{S} = \{s^{(i)}\}_{i=1}^{m}$ denote a set of sessions over an item set $\mathcal{I} = \{i_1, \ldots, i_n\}$. An anonymous session $s \in \mathcal{S}$ is a sequence of items $s = (s_1, s_2, \dots, s_{|s|})$, where $s_j \in \mathcal{I}$ is the $j$-th clicked item and $|s|$ is the length of the session $s$. Given $s \in \mathcal{S}$, we represent a training example $\mathbf{x} \in \mathbb{R}^{n}$, where each element $\mathbf{x}_j$ is non-zero if the $j$-{th} item in $\mathcal{I}$ is observed in the session and zero otherwise. Higher values are assigned if the interacted item is more important. Similarly, the target item is represented as a vector $\mathbf{y} \in \mathbb{R}^{n}$, where $y_j = 1$ if $i_j\in s_{|s|+1}$, and 0 otherwise.

Stacking $m$ examples, we denote $\mathbf{X}, \mathbf{Y} \in \mathbb{R}^{m \times n}$ for a session-item interaction matrix for training examples and target labels, respectively, where $m$ is the number of training examples. The session-item interaction matrix is extremely sparse by nature. Generally speaking, $m$ can be an arbitrary number depending on how we construct the examples and how much we sample from the data. (See Section~\ref{sec:session_aware_models} for more details.)

\vspace{1mm}
\noindent
\textbf{Problem statement.} The goal of session-based recommendation is to predict the next item(s) a user would likely choose to consume, given a sequence of previously consumed items in a session. Formally, we build a session-based model $\mathbf{M}(s)$ that takes a session $s = (s_1, s_2, \dots, s_t)$ for $t = 1, 2, ..., |s|-1$ as input and returns a list of top-$N$ candidate items to be consumed as the next one $s_{t+1}$.

\vspace{1mm}
\noindent
\textbf{Linear item-item models.} Given two matrices $\mathbf{X}$ and $\mathbf{Y}$, the linear model is formulated with an item-to-item similarity matrix $\mathbf{B} \in \mathbb{R}^{n \times n}$. Formally, a linear item-item model is defined by
\begin{equation}
  \mathbf{Y} = \mathbf{X} \cdot \mathbf{B},
\end{equation}
where $\mathbf{B}$ maps the previously consumed items in $\mathbf{X}$ to the next observed item(s) in $\mathbf{Y}$. A typical objective function of this linear model is formulated by ridge regression that minimizes the ordinary least squares (OLS).
\begin{equation}\label{eq:linear}
  \argmin_{\mathbf{B}} \|\mathbf{Y} - \mathbf{X} \cdot \mathbf{B}\|_F^2 + \lambda  \|\mathbf{B}\|_F^2,
\end{equation}
where $\lambda$ is a regularizer term and $\|\cdot\|_F$ denotes the Frobenius norm.

In the traditional recommendation, where each user is represented as a set of all items consumed without the concept of sessions, $\mathbf{X}$ and $\mathbf{Y}$ are regarded as the same matrix. In this case, with $\lambda = 0$, it ends up with a trivial solution of $\mathbf{B}=\mathbf{I}$ in Eq.~\eqref{eq:linear}, which is useless for predictions. To avoid this trivial solution, existing studies~\cite{NingK11, Steck19b} add some constraints to the objective function. As the pioneering work, SLIM~\cite{NingK11} enforces all entries in $\mathbf{B}$ to be non-negative with zero diagonal elements.
\begin{equation}
  \label{eq:slim_loss}
  \begin{aligned}
    \argmin_{\mathbf{B}} & \|\mathbf{X} - \mathbf{X} \cdot \mathbf{B}\|_F^2 + \lambda_1  \|\mathbf{B}\|_1 + \lambda_2  \|\mathbf{B}\|_F^2 \\
    & \text{s.t.} \ \ \texttt{diag}(\mathbf{B}) = 0, \ \ \mathbf{B} \ge 0,
  \end{aligned}
\end{equation}
where $\lambda_1$ and $\lambda_2$ are regularization coefficients for L1-norm and L2-norm, respectively, and $\texttt{diag}(\mathbf{B}) \in \mathbb{R}^{n}$ denotes a vector with the diagonal elements of $\mathbf{B}$.

Although SLIM~\cite{NingK11} has shown competitive accuracy in literature, it is also well-known SLIM is prohibitively slow to train. (\citet{LiangKHJ18} reported that the hyper-parameter tuning of SLIM took several weeks on large-scale datasets with 10K+ items.) Although some extensions~\cite{SedhainMSB16, SedhainBKVKMBS16,  SteckDRJ20} propose to reduce the training cost, they are still computationally prohibitive at an industrial scale.

Recently, EASE$^{R}$~\cite{Steck19b} and its variants~\cite{Steck19a,Steck19c} remove the non-negativity constraint of $\mathbf{B}$ and L1-norm constraints from Eq.~\eqref{eq:slim_loss}, leaving only the diagonal constraint.
\begin{equation}
  \label{eq:ease_loss}
  \argmin_{\mathbf{B}} \|\mathbf{X} - \mathbf{X} \cdot \mathbf{B}\|_F^2 + \lambda \cdot \|\mathbf{B}\|_F^2 \ \ \ \text{s.t.} \ \ \texttt{diag}(\mathbf{B}) = 0.
\end{equation}
EASE$^{R}$ draws the closed-form solution via Lagrange multipliers.
\begin{equation}
  \label{eq:ease_solution}
  \hat{\mathbf{B}} = \mathbf{I} - \hat{\mathbf{P}} \cdot  \texttt{diagMat}(\textbf{1} \oslash \texttt{diag}(\hat{\mathbf{P}})),
\end{equation}
where $\hat{\mathbf{P}} = (\mathbf{X}^{\top} \mathbf{X} + \lambda \mathbf{I})^{-1}$, $\textbf{1}$ is a vector of ones, and $\oslash$ denotes the element-wise division operator. Also, $\texttt{diagMat}(\mathbf{x})$ denote the diagonal matrix expanded from vector $\mathbf{x}$. (See~\cite{Steck19b} for a full derivation of the closed-form solution.) Finally, each learned weight in $\mathbf{B}$ is given by
\begin{equation}
  \hat{\mathbf{B}}_{ij} = 
  \begin{cases}
    \ \ \ 0 & \text{if} \ \ i = j, \\
    \ \ - \frac{\mathbf{P}_{ij}}{\mathbf{P}_{jj}} & \text{otherwise}.
  \end{cases}
\end{equation}

Although inverting the regularized Gram matrix is a bottleneck for a large-scale dataset, the closed-form expression is significantly advantageous in efficiency. The complexity of EASE$^{R}$~\cite{Steck19b} is proportional to the number of items, which is usually much smaller than the number of users in real-world scenarios. It also achieves competitive prediction accuracy to the state-of-the-art models in the conventional recommendation setting. Motivated by these advantages, we utilize the benefit of linear models for session-based recommendation.

\section{Proposed Models}
\label{sec:model}

\subsection{Motivation}
\label{sec:motivation}

We present the useful characteristics of sessions. The first three are about the correlation between items in a session, \ie, \emph{intra-session} properties, while the last one is about the relationship across sessions, \ie, \emph{inter-session} property.

\begin{itemize}[leftmargin=5mm]
  \item \textbf{Session consistency}: A list of items in a session is often topically coherent, reflecting a specific and consistent user intent. For example, the songs in a user-generated playlist usually have a theme, \eg, similar mood, same genre, or artist.
  
  \item \textbf{Sequential dependency}: Some items tend to be observed in a particular order. An item is usually consumed first, followed by another, across the majority of sessions. An intuitive example is a TV series, where its episodes are usually sequentially watched. For this reason, the last item in the current session is often the most informative signal to predict the next one.
  
  \item \textbf{Repeated item consumption}: A user may repeatedly consume an item multiple times in a single session. For instance, a user may listen to her favorite songs repeatedly. Also, a user may click the same product multiple times to compare it with other products.
  
  \item \textbf{Timeliness of sessions}: A session usually reflects user interest at the moment, and a collection of such sessions often reflects a recent trend. For instance, many users tend to watch newly released music videos more frequently. Thus, the recent sessions may be a better indicator of predicting the user's current interest than the past ones.
\end{itemize}

Recent session-based recommendation models have implicitly utilized some of these properties. RNN-based~\cite{HidasiKBT15, HidasiQKT16, HidasiK18} or attention-based models~\cite{LiRCRLM17, LiuZMZ18} assume that the last item in the session is the most crucial user intent and mainly focus on \emph{sequential dependency}. GNN-based models~\cite{WuT0WXT19,XuZLSXZFZ19,GuptaGMVS19,PanCLR20a} can also utilize \emph{session consistency}. They leverage the graph embeddings between items to analyze hidden user intents. Some studies also addressed \emph{repeated item consumption} by using the attention mechanism (\eg, RepeatNet~\cite{RenCLR0R19}) or \emph{timeliness of sessions} by using the weighting scheme (\eg, STAN~\cite{GargGMVS19}), thereby improving the prediction accuracy.

However, none of these existing studies holistically deals with the various session properties. Each model is optimized to handle a few properties, overlooking other essential characteristics. Also, different datasets exhibit quite diverse tendencies. For instance, the Yoochoose dataset tends to show stronger sequential dependency than other properties, while the sessions in the Diginetica dataset rely less on sequential dependency. As a result, no single model outperforms others across multiple datasets~\cite{LudewigJ18, LudewigMLJ19a, LudewigMLJ19b}, failing to generalize on them. Besides, existing DNN-based models are inapplicable to resource-limited environments due to the lack of scalability. To overcome these limitations, we develop session-aware linear models that not only simultaneously accommodate various properties of sessions but also are highly scalable.

\subsection{Session Representations}
\label{sec:session_representations}

\begin{figure}
\includegraphics[width=1.0\linewidth]{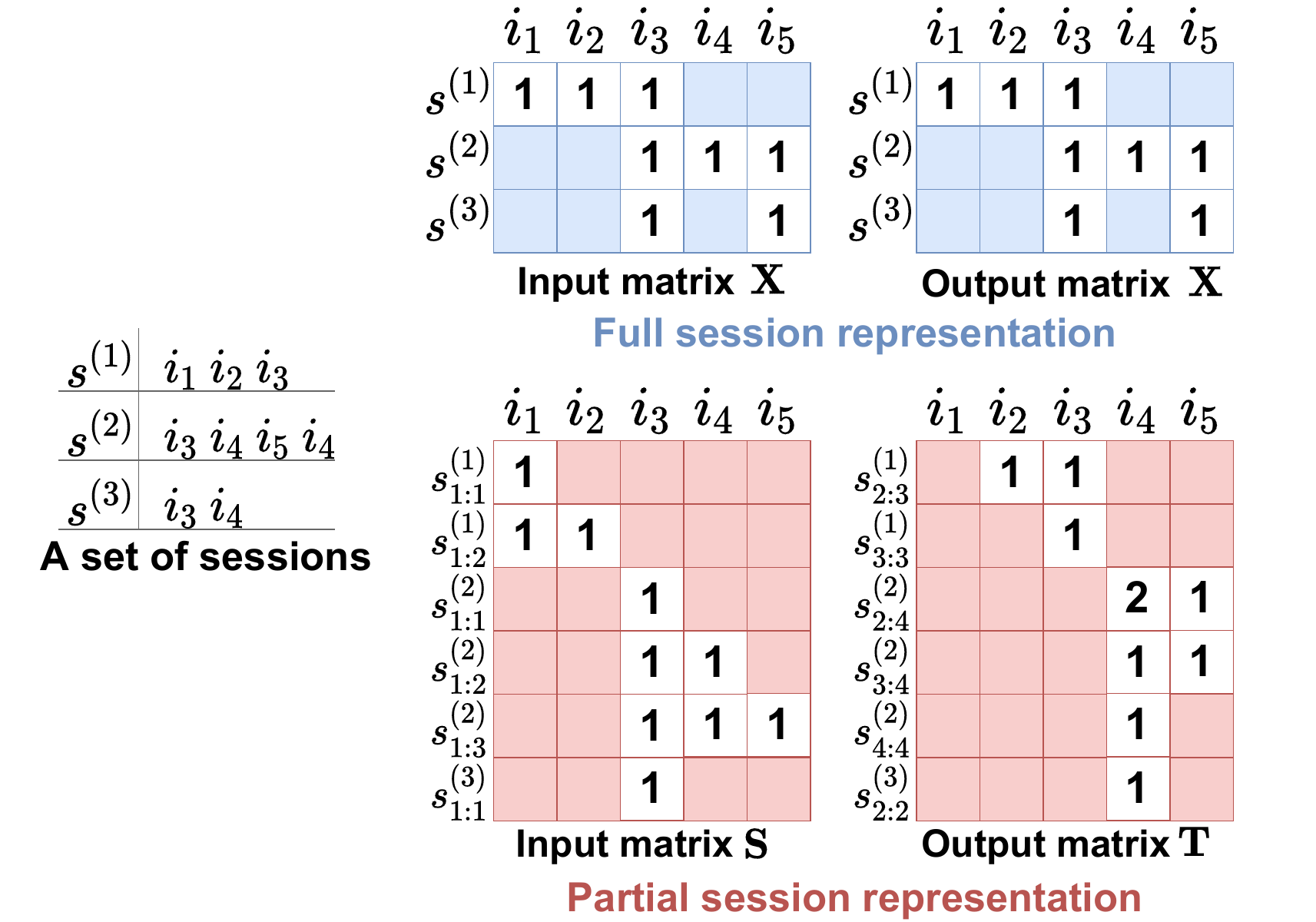}
\caption{An example of two session representations. In full session representation, all items in a session are related each other. In partial session representation, there are sequential correlations across items.}\label{fig:session_representation}
\end{figure}

We describe how to represent the sessions for our linear models. Although the session has a sequential nature, the linear model only handles a set of items as a vector. As depicted in Figure~\ref{fig:session_representation}, we consider two session representations.
\begin{itemize}[leftmargin=5mm]
  \item \textbf{Full session representation}: The entire session is represented as a single vector. Specifically, a session $s = (s_1, s_2, \ldots, s_{|s|})$ is regarded as a set of items $\mathbf{s} = \{s_1, s_2, \ldots, s_{|s|}\}$, ignoring the sequence of items. As depicted in Figure~\ref{fig:session_representation}, the number of training examples ($m$) is equal to the number of sessions. It is more suitable for a case where items in a session tend to have a stronger correlation among them, relatively insensitive to the order of consumption. Note that the repeated items in the session are treated as a single item as the full session representation mainly handles the session consistency across items.

  \item \textbf{Partial session representation}: A session is divided into two subsets, \emph{past} and \emph{future}, to represent the sequential correlations across items. For some time step $t$, the past set consists of items that have been consumed before $t$, \ie, $\mathbf{s}_{1:t-1} = \{s_1, \ldots, s_{t-1}\}$, and the future set consists of items consumed at or after $t$, \ie, $\mathbf{s}_{t:|s|} = \{s_t, \dots, s_{|s|}\}$. We can produce $|s| - 1$ such partial sessions by taking each $t = 2, ..., |s|$ as the threshold~\cite{GargGMVS19}. By stacking $|s| - 1$ past and future set pairs for all sessions, we construct two matrices, $\mathbf{S}, \mathbf{T} \in \mathbb{R}^{m' \times n}$, where $m'$ is the number of all partial sessions extended from $\mathcal{S}$, \ie, $m' = \sum_{i=1}^{m}{(|s^{(i)}| - 1})$. Using $\mathbf{S}$ and $\mathbf{T}$, we learn sequential dependency across items.
\end{itemize}

\subsection{Session-aware Linear Models}
\label{sec:session_aware_models}

First, we devise two linear models that utilize full and partial session representations, respectively. Then, we unify two linear models to fully capture the multifaceted aspects of the sessions.

\subsubsection{Session-aware Linear Item Similarity Model (SLIS)}
\label{sec:slis}

We present a linear model using full session representation, focusing on the similarity between items. As described in Section~\ref{sec:session_representations}, the input and output matrix ($\mathbf{X}$) are same, as handled in existing linear models~\cite{NingK11,Steck19b}, \ie, $\mathbf{X} = \mathbf{X} \cdot \mathbf{B}$. However, the existing models do not handle various characteristics of sessions other than session consistency, leaving them sub-optimal.

We propose a new linear model by reformulating the objective function of SLIM~\cite{NingK11} to accommodate the timeliness of sessions and repeated item consumption in the session. Firstly, we adopt a weight matrix $\mathbf{W} \in \mathbb{R}^{m \times n}$ that corresponds to $\|\mathbf{X} - \mathbf{X} \cdot \mathbf{B}\|_F^2$. Assuming the timeliness of sessions decays over time, $\mathbf{W}$ is used to distinguish the timeliness of sessions. Secondly, we relax the zero-diagonal constraint for $\mathbf{B}$ to handle repeated item consumption. Since the diagonal element of $\mathbf{B}$ is loosely penalized, our model allows us to predict the same item as the next item repeatedly. Formally, the objective function of SLIS is formulated by
\begin{equation}
  \label{eq:slis_loss}
  \argmin_{\mathbf{B}} \left\|\mathbf{W} \odot ( \mathbf{X} - \mathbf{X} \cdot \mathbf{B} )\right\|_F^2 + \lambda \|\mathbf{B}\|_F^2,
\ \ s.t. \ \ \texttt{diag}(\mathbf{B}) \le \xi,
\end{equation}
where $\odot$ denotes the element-wise product operator, and $\xi$ is a hyperparameter to control the diagonal constraint. When $\xi = 0$, it is equivalent to the zero-diagonal constraint for $\mathbf{B}$. When $\xi = \infty$, there is no constraint for the diagonal elements of $\mathbf{B}$. Note that the objective function in Eq.~\eqref{eq:slis_loss} is the generalized version of the objective function of EASE$^{R}$~\cite{Steck19b} in Eq.~\eqref{eq:ease_loss}.

The key advantage of EASE$^{R}$~\cite{Steck19b} is its closed-form solution. Can we still solve our objective function in Eq.~\eqref{eq:slis_loss} by a  closed-form solution, despite these modifications? We can still achieve a closed-form solution for the relaxed diagonal constraint without $\mathbf{W}$ via Karush–Kuhn–Tucker (KKT) conditions. For an arbitrary weight matrix $\mathbf{W}$, however, it is non-trivial to obtain a closed-form solution.

To address this issue, we deal with two special cases for $\mathbf{W}$, \emph{weights of sessions} and \emph{weights of items}. As presented in~\cite{Steck19c}, the weights of items do not affect learning $\mathbf{B}$. Thus, we only consider the weights of sessions; $\mathbf{W}$ is treated as the weight vector of sessions to distinguish the importance of sessions. Let $\mathbf{w}_\text{full}$ denote the weight vector for each session in $\mathcal{S}$. Then, $\mathbf{W}$ is replaced by the outer product of the session weight and the one vector, \ie, $\mathbf{W}_\text{full} = \mathbf{w}_\text{full} \cdot \textbf{1}^{\top}$, where $\mathbf{w}_{\text{full}} \in \mathbb{R}^{m}$ and $\textbf{1} \in \mathbb{R}^{n}$. The linear model with $\mathbf{W}_\text{full}$ is solved by the closed-form equation.
\begin{equation}
  \label{eq:slis_solution}
  \begin{aligned}
    \hat{\mathbf{B}} = & \ \mathbf{I} -  \hat{\mathbf{P}} \cdot  \texttt{diagMat}(\mathbf{\gamma}), \\
    \text{where} \ \ \mathbf{\gamma}_{j} = &
    \begin{cases}
        \ \ \ \lambda & \text{if} \ \ \ 1 - \lambda \mathbf{P}_{jj} \le \xi, \\
        \ \ \ \frac{1-\xi}{P_{jj}} & \text{otherwise}. 
    \end{cases}
\end{aligned}
\end{equation}
Note that $\hat{\mathbf{P}} = (\mathbf{X}^{\top} \cdot \texttt{diagMat}(\mathbf{w}^{2}_\text{full}) \cdot \mathbf{X} + \lambda \mathbf{I})^{-1}$. $\mathbf{\gamma} \in \mathbb{R}^n$ is the vector to check the diagonal constraint of $\mathbf{B}$. (See the detailed derivation of closed-form solution of our linear models in Appendix~\ref{sec:app_solution}.) Because of the inequality constraint for diagonal elements in $\mathbf{B}$, $\mathbf{\gamma}_{j}$ is determined by the condition of $(1 - \lambda \mathbf{P}_{jj})$. When the $j$-th constraint $\mathbf{B}_{jj}\leq \xi$ is violated, $\mathbf{\gamma}_{j}$ regularizes $\mathbf{B}$ to satisfy the constraint. Depending on $\xi$, the closed-form solution is different; when $\xi = 0$, $\mathbf{\gamma}_{j}$ is equal to ${1} / \mathbf{P}_{jj}$, corresponding to $(1 \oslash \texttt{diag}(\hat{\mathbf{P}}))$. When $\xi = \infty$, the closed-form solution is $\hat{\mathbf{B}}=\mathbf{I}-\lambda\hat{\mathbf{P}}$.

Lastly, we describe how to set the weight of sessions for $\mathbf{W}$. As discussed in~\cite{GargGMVS19}, we assume the significance of sessions decays over time. To reflect the timeliness of sessions, we assign higher weights to more recent sessions by
\begin{equation}
  \label{eq:session_decay}
  \mathbf{w}_\text{time}(s) = \text{exp}\left(-\frac{t_{ \text{max}} - t(s)}{\delta_\text{time}} \right),
\end{equation}
where $\delta_\text{time}$ is the hyperparameter to control the decay rate, $t_{\text{max}}$ is the most recent timestamp in $\mathcal{S}$, and $t(s)$ is the timestamp of the session $s$. As $\delta_\text{time}$ is small, we decay the significance of old sessions. Finally, we assign the $j$-th value of $\mathbf{w}_\text{full}$ as $\sqrt{\mathbf{w}_\text{time}(j)}$.

\subsubsection{Session-aware Linear Item Transition Model (SLIT)}
\label{sec:slit}

Using the partial session representation in Section~\ref{sec:session_representations}, we design a linear model that captures the sequential dependency across items. Unlike SLIS, each session is split into multiple partial sessions, forming different input and output matrices. Similar to SLIS, we also incorporate the weight of sessions to SLIT. Meanwhile, we ignore the constraint for diagonal elements in $\mathbf{B}$ as different input and output matrices are naturally free from the trivial solution.

Formally, we formulate the objective function of SLIT using the partial session representation.
\begin{equation}
  \label{eq:slit_loss}
  \argmin_{\mathbf{B}} \left\|\mathbf{W}_\text{par} \odot ( \mathbf{T} - \mathbf{S} \cdot \mathbf{B} )\right\|_F^2 + \lambda \|\mathbf{B}\|_F^2,
\end{equation}
where $\mathbf{S}, \mathbf{T} \in \mathbb{R}^{m' \times n}$ denote the past and future matrices. Assuming the partial sessions have the timestamp of the original session, $\mathbf{W}_\text{par}$ is represented by $\mathbf{w}_\text{par} \cdot \textbf{1}^{\top}$. Similar to $\mathbf{w}_\text{full}$, the values in $\mathbf{w}_\text{par}$ are assigned by Eq.~\eqref{eq:session_decay}. Although $\mathbf{S}$ and $\mathbf{T}$ are different, we still derive the closed-form solution:
\begin{equation}
  \label{eq:slit_solution}
  \hat{\mathbf{B}} = \hat{\mathbf{P}}' \cdot [\mathbf{S}^{\top} \texttt{diagMat}(\mathbf{w}^{2}_\text{par}) \mathbf{T}],
\end{equation}
where $\hat{\mathbf{P}}' = (\mathbf{S}^{\top} \texttt{diagMat}(\mathbf{w}^{2}_\text{par}) \mathbf{S} + \lambda \mathbf{I})^{-1}$.

Lastly, we describe how to adjust the importance of the past and future item subsets in $\mathbf{S}$ and $\mathbf{T}$. As adopted in~\cite{GargGMVS19}, we utilize the position gap between two items as the weight of items.
\begin{equation}
  \label{eq:weight_item}
  \mathbf{w}_\text{pos}(i, j, s) = \text{exp}\left(-\frac{|p(i, s) - p(j, s)|}{\delta_\text{pos}} \right),
\end{equation}
where $\delta_\text{pos}$ is the hyperparameter to control the position decay in partial sessions, and $p(i, s)$ is the position of item $i$ in session $s$. Also, $i$ and $j$ are the items in session $s$. When a target item is given, it is possible to decay the importance of items in the partial session according to sequential order.

\subsubsection{Unifying Two Linear Models}

Since SLIS and SLIT capture various characteristics of sessions, we propose a unified model, called \emph{Session-aware Linear Similarity/Transition model (SLIST)}, by jointly optimizing both models:

\begin{equation}
  \label{eq:joint_loss1}
  \begin{aligned}
    \argmin_{\mathbf{B}} & \;\alpha\left\|\mathbf{W}_\text{full} \odot ( \mathbf{X} - \mathbf{X} \cdot \mathbf{B} )\right\|_F^2 \\
    & + (1-\alpha) \left\|\mathbf{W}_\text{par} \odot ( \mathbf{T} - \textbf{S} \cdot \mathbf{B} )\right\|_F^2   +  \lambda \|\mathbf{B}\|_F^2 , \\
  \end{aligned}
\end{equation}
where $\alpha$ is the hyperparameter to control the importance ratio of SLIS and SLIT.

Although it looks impossible to achieve a closed-form solution at a glance, we can still derive it. The intuition is that, when stacking two matrices for SLIS and SLIT, \eg, $\mathbf{X}$ and $\mathbf{S}$, the objective function of SLIST is similar to that of SLIT. Formally, the closed-form solution is given by
\begin{equation}
  \hat{\mathbf{B}} = \mathbf{I} - \lambda\hat{\mathbf{P}} - (1-\alpha)\hat{\mathbf{P}} \mathbf{S}^{\top}\texttt{diagMat}(\mathbf{w}^{2}_\text{par})(\mathbf{S}-\mathbf{T}),
\end{equation}
where $\hat{\mathbf{P}} = (\alpha\mathbf{X}^{\top}\texttt{diagMat}(\mathbf{w}^{2}_\text{full})\mathbf{X} + (1-\alpha) \mathbf{S}^{\top} \texttt{diagMat}(\mathbf{w}^{2}_\text{par}) \mathbf{S} + \lambda\mathbf{I})^{-1}$.

\subsubsection{Model Inference}
We predict the next item for a given session with a sequence of items using a learned item-item matrix, \ie, $\hat{\mathbf{T}} = \mathbf{S} \cdot \hat{\mathbf{B}}$. For inference, it is necessary to consider the importance of items depending on a sequence of items in $\mathbf{S}$. Similar to Eq.~\eqref{eq:weight_item}, we decay the item weights for inference over time by
\begin{equation}
  \label{eq:weight_inference}
  \mathbf{w}_\text{inf}(i, s) = \text{exp}\left(-\frac{|s| - p(i, s)}{\delta_\text{inf}} \right),
\end{equation}
where $\delta_\text{inf}$ is the hyperparameter to control the item weight decay for inference, and $|s|$ is the length of the session $s$. In other words, SLIST takes the vector by partial representation as input and use it with the item weight.
\section{Experimental Setup}
\label{sec:setup}

\textbf{Benchmark datasets}. We evaluate our proposed models on four public datasets collected from e-commerce and music streaming services: YooChoose\footnote{https://2015.recsyschallenge.com/challenge.html} (YC), Diginetica\footnote{http://cikm2016.cs.iupui.edu/cikm-cup} (DIGI), RetailRocket\footnote{https://www.dropbox.com/sh/n281js5mgsvao6s/AADQbYxSFVPCun5DfwtsSxeda?dl=0} (RR), and NowPlaying (NOWP). For YC and DIGI, we use single-split datasets (\ie, YC-1/64, YC-1/4, DIGI1), following existing studies~\cite{HidasiK18,LiRCRLM17,LiuZMZ18,WuT0WXT19}. To evaluate on large-scale datasets, we further employ five-split datasets (\ie, YC, DIGI5, RR, NOWP), used in recent empirical analysis~\cite{LudewigJ18,LudewigMLJ19a,LudewigMLJ19b} for session-based recommendation models. Table~\ref{tab:statistics} summarizes detailed statistics of all benchmark datasets.

As pre-processing, we discard the sessions having only one interaction and items appearing less than five times following the convention~\cite{LudewigJ18}. We hold-out the sessions from the last $N$-days for test purposes and used the last $N$ days in the training set for the validation set. Refer to detailed pre-processing and all source codes at our website\footnote{https://github.com/jin530/SLIST}.

\vspace{1mm}
\noindent
\textbf{Competing models}. We compare our models with eight competitive models, roughly categorized into four groups, Markov-chain-based, neighborhood-based, linear, and DNN models. SR~\cite{KamehkhoshJL17} is an effective Markov chain model together with association rules. Among neighborhood-based models, we choose SKNN~\cite{JannachL17} and STAN~\cite{GargGMVS19}. STAN~\cite{GargGMVS19} is an improved version of SKNN~\cite{JannachL17} by considering sequential dependency and session recency. For the linear model baselines, we apply EASE$^{R}$~\cite{Steck19b} for the session-based recommendation problem by using the session-item matrix, namely SEASE$^{R}$. For DNN-based models, we employ GRU4Rec$^+$~\cite{HidasiKBT15,HidasiK18}, NARM~\cite{LiRCRLM17}, STAMP~\cite{LiuZMZ18}, and SR-GNN~\cite{WuT0WXT19}. GRU4Rec$^+$~\cite{HidasiK18} is an improved version of the original GRU4Rec\cite{HidasiQKT16} using the top-$k$ gain objective function. NARM~\cite{LiRCRLM17} and STAMP~\cite{LiuZMZ18} are attention-based models for facilitating both long-term and short-term sequential dependency. SR-GNN~\cite{WuT0WXT19} utilizes a graph neural network to capture complex sequential dependency within a session.

\vspace{1mm}
\noindent
\textbf{Evaluation protocol}. To evaluate session-based recommender models, we adopt the \emph{iterative revealing scheme}~\cite{HidasiQKT16, LiRCRLM17, WuT0WXT19}, which iteratively exposes the item of a session to the model. Each item in the session is sequentially appended to the input of the model. Therefore, this scheme is useful for reflecting the sequential user behavior throughout a session.

\begin{table}
\small
\caption{Detailed statistics of the benchmark datasets. \#Actions indicates the number of entire user-item interactions.}
\label{tab:statistics}
\begin{tabular}{ll|rrrrr}
\toprule
Split & Dataset & \#Actions & \#Sessions & \#Items & \#Actions & \#Items \\
& & & & & / Sess. & / Sess.\\
\hline
1-split
& YC-1/64 & 494,330   & 119,287    & 17,319  & 4.14          & 3.30         \\
& YC-1/4  & 7,909,307 & 1,939,891  & 30,638  & 4.08          & 3.28        \\
& DIGI1  & 916,370   & 188,807    & 43,105  & 4.85          & 4.08        \\
\hline
5-split
& YC      & 5,426,961 & 1,375,128  & 28,582  & 3.95          & 3.17        \\
& DIGI5  & 203,488   & 41,755     & 32,137  & 4.86          & 4.08        \\
& RR      & 212,182   & 59,962     & 31,968  & 3.54          & 2.56        \\
& NOWP    & 271,177   & 27,005     & 75,169  & 10.04         & 9.38        \\
\bottomrule
\end{tabular}
\end{table}

\begin{table*}[t] \small
\caption{Accuracy comparison of our models and competing models, following experimental setup in \cite{LudewigMLJ19a, LudewigMLJ19b}. Gains indicate how much better the best proposed model is than the best competing model. The best proposed model is marked in \com{\textbf{bold}} and the best baseline model is \fcom{\underline{underlined}}.}
\label{tab:overallresult}
\begin{center}
\begin{tabular}{c|c|cccc|cccc|ccc|c}
\toprule
\multirow{2}{*}{Dataset} & \multirow{2}{*}{Metric} & \multicolumn{4}{c|}{Non-DNN-based models}         & \multicolumn{4}{c|}{DNN-based models}                  & \multicolumn{3}{c|}{Ours}                & \multirow{2}{*}{Gains (\%)} \\
                         &                         & SR     & SKNN         & STAN         & SEASE$^{R}$ & GRU4Rec$^+$      & NARM         & STAMP  & SR-GNN       & SLIS             & SLIT             & SLIST          &                           \\
\midrule
\multirow{4}{*}{YC-1/64} & HR@20                   & 0.6674 & 0.6423       & 0.6838       & 0.5443      & 0.6528       & 0.6998       & 0.6841 & \fcom{\ul{ 0.7021}} & 0.7015          & 0.6968          & \com{\textbf{0.7088}} & 0.95                        \\
                         & MRR@20                  & 0.3033 & 0.2522       & 0.2820       & 0.1963      & 0.2752       & 0.2957       & 0.2995 & \fcom{\ul{ 0.3099}} & 0.2837          & \com{\textbf{0.3093}} & 0.3083          & -0.19                       \\
                         & R@20                    & 0.4709 & 0.4780       & 0.4955       & 0.3983      & 0.4009       & \fcom{\ul{ 0.5051}} & 0.4904 & 0.5049       & 0.5051          & 0.4976          & \com{\textbf{0.5080}} & 0.57                        \\
                         & MAP@20                  & 0.0350 & 0.0363       & 0.0375       & 0.0306      & 0.0285       & \fcom{\ul{ 0.0389}} & 0.0370 & 0.0388       & 0.0387          & 0.0379          & \com{\textbf{0.0388}} & -0.26                       \\ \hline
\multirow{4}{*}{YC-1/4}  & HR@20                   & 0.6850 & 0.6329       & 0.6846       & 0.5550      & 0.6940       & 0.7079       & 0.7021 & \fcom{\ul{ 0.7118}} & 0.7119          & 0.7149          & \com{\textbf{0.7175}} & 0.80                        \\
                         & MRR@20                  & 0.3053 & 0.2496       & 0.2829       & 0.1984      & 0.2942       & 0.2996       & 0.3066 & \fcom{\ul{ 0.3180}} & 0.283           & 0.3155          & \com{\textbf{0.3161}} & -0.60                       \\
                         & R@20                    & 0.4851 & 0.4756       & 0.4952       & 0.4040      & 0.4887       & \fcom{\ul{ 0.5097}} & 0.5008 & 0.5095       & 0.5121          & 0.511           & \com{\textbf{0.5130}} & 0.65                        \\
                         & MAP@20                  & 0.0364 & 0.0361       & 0.0373       & 0.0309      & 0.0375       & \fcom{\ul{ 0.0395}} & 0.0385 & 0.0393       & \com{\textbf{0.0394}} & 0.0391          & 0.0393          & -0.25                       \\ \hline
\multirow{4}{*}{DIGI1}   & HR@20                   & 0.4085 & 0.4846       & \fcom{\ul{ 0.5121}} & 0.3906      & 0.5097       & 0.4979       & 0.4690 & 0.4904       & 0.5247          & 0.4729          & \com{\textbf{0.5291}} & 3.32                        \\
                         & MRR@20                  & 0.1431 & 0.1736       & \fcom{\ul{ 0.1851}} & 0.1099      & 0.1750       & 0.1585       & 0.1499 & 0.1654       & 0.1877          & 0.1599          & \com{\textbf{0.1886}} & 1.89                        \\
                         & R@20                    & 0.3164 & 0.3788       & \fcom{\ul{ 0.3965}} & 0.3048      & 0.3957       & 0.3890       & 0.3663 & 0.3811       & 0.4062          & 0.3651          & \com{\textbf{0.4091}} & 3.18                        \\
                         & MAP@20                  & 0.0212 & 0.0263       & \fcom{\ul{ 0.0275}} & 0.0209      & \fcom{\ul{ 0.0275}} & 0.0270       & 0.0252 & 0.0264       & 0.0284          & 0.0250          & \com{\textbf{0.0286}} & 4.00                        \\ \midrule
\multirow{4}{*}{YC}      & HR@20                   & 0.6506 & 0.5996       & 0.6656       & 0.5030      & 0.6488       & \fcom{\ul{ 0.6751}} & 0.6654 & 0.6713       & 0.6786          & 0.6838          & \com{\textbf{0.6867}} & 1.72                        \\
                         & MRR@20                  & 0.3010 & 0.2620       & 0.2933       & 0.1849      & 0.2893       & 0.3047       & 0.3033 & \fcom{\ul{ 0.3142}} & 0.2854          & 0.3080          & \com{\textbf{0.3097}} & -1.43                       \\
                         & R@20                    & 0.4853 & 0.4658       & 0.4986       & 0.3809      & 0.4837       & \fcom{\ul{ 0.5109}} & 0.4979 & 0.5060       & 0.5074          & 0.5097          & \com{\textbf{0.5122}} & 0.25                        \\
                         & MAP@20                  & 0.0332 & 0.0318       & 0.0342       & 0.0264      & 0.0334       & \fcom{\ul{ 0.0357}} & 0.0344 & 0.0351       & 0.0353          & 0.0355          & \com{\textbf{0.0357}} & 0.00                        \\ \hline
\multirow{4}{*}{DIGI5}   & HR@20                   & 0.3277 & 0.4748       & \fcom{\ul{ 0.4800}} & 0.3531      & 0.4639       & 0.4188       & 0.3917 & 0.4158       & 0.4939          & 0.4193          & \com{\textbf{0.5005}} & 4.27                        \\
                         & MRR@20                  & 0.1216 & 0.1714       & \fcom{\ul{ 0.1828}} & 0.1004      & 0.1644       & 0.1392       & 0.1314 & 0.1436       & \com{\textbf{0.1847}} & 0.1400          & 0.1827          & 1.04                        \\
                         & R@20                    & 0.2517 & 0.3715       & \fcom{\ul{ 0.3720}} & 0.2774      & 0.3617       & 0.3254       & 0.3040 & 0.3232       & 0.3867          & 0.3260          & \com{\textbf{0.3898}} & 4.78                        \\
                         & MAP@20                  & 0.0164 & \fcom{\ul{ 0.0255}} & 0.0252       & 0.0185      & 0.0247       & 0.0218       & 0.0201 & 0.0217       & 0.0265          & 0.0218          & \com{\textbf{0.0267}} & 4.71                        \\ \hline
\multirow{4}{*}{RR}      & HR@20                   & 0.4174 & 0.5788       & \fcom{\ul{ 0.5938}} & 0.2727      & 0.5669       & 0.5549       & 0.4620 & 0.5433       & 0.5983          & 0.4899          & \com{\textbf{0.6020}} & 1.38                        \\
                         & MRR@20                  & 0.2453 & 0.3370       & \fcom{\ul{ 0.3638}} & 0.1032      & 0.3237       & 0.3196       & 0.2527 & 0.3066       & \com{\textbf{0.3583}} & 0.2691          & 0.3512          & -1.51                       \\
                         & R@20                    & 0.3359 & 0.4707       & \fcom{\ul{ 0.4748}} & 0.2209      & 0.4559       & 0.4526       & 0.3917 & 0.4438       & 0.4798          & 0.3925          & \com{\textbf{0.4819}} & 1.50                        \\
                         & MAP@20                  & 0.0194 & 0.0283       & \fcom{\ul{ 0.0285}} & 0.0134      & 0.0272       & 0.0270       & 0.0227 & 0.0264       & 0.0289          & 0.0233          & \com{\textbf{0.0291}} & 2.11                        \\ \hline
\multirow{4}{*}{NOWP}    & HR@20                   & 0.2002 & \fcom{\ul{ 0.2450}} & 0.2414       & 0.2088      & 0.2261       & 0.1849       & 0.1915 & 0.2113       & 0.2522          & 0.2326          & \com{\textbf{0.2689}} & 9.76                        \\
                         & MRR@20                  & 0.1052 & 0.0687       & 0.0871       & 0.0625      & \fcom{\ul{ 0.1076}} & 0.0894       & 0.0882 & 0.0935       & 0.0794          & \com{\textbf{0.1171}} & 0.1137          & 8.83                        \\
                         & R@20                    & 0.1366 & \fcom{\ul{ 0.1809}} & 0.1696       & 0.1658      & 0.1361       & 0.1274       & 0.1253 & 0.1400       & 0.1802          & 0.1577          & \com{\textbf{0.1840}} & 1.71                        \\
                         & MAP@20                  & 0.0133 & \fcom{\ul{ 0.0186}} & 0.0175       & 0.0181      & 0.0116       & 0.0118       & 0.0113 & 0.0125       & 0.0183          & 0.0146          & \com{\textbf{0.0184}} & -1.08\\ 
\bottomrule
\end{tabular}
\end{center}
\end{table*}

\vspace{1mm}
\noindent
\textbf{Evaluation metrics}. We utilize various evaluation metrics with the following two application scenarios, \eg, a set of recommended products in e-commerce and the next recommended song for a given playlist in music streaming services. (i) To predict the next item in a session, we use \emph{Hit Rate (HR)} and \emph{Mean Reciprocal Rank (MRR)}, which have been widely used in existing studies~\cite{HidasiKBT15, LiuZMZ18, WuT0WXT19}. (ii) To consider all subsequent items for the session, we use \emph{Recall} and \emph{Mean Average Precision (MAP)} as the standard IR metrics.

\vspace{1mm}
\noindent
\textbf{Implementation details}. For SLIST, we tune $\alpha$ among \{0.2, 0.4, 0.6, 0.8\},  $\delta_\text{pos}$ and $\delta_\text{inf}$ among \{0.125, 0.25, 0.5, 1, 2, 4, 8\} and $\delta_\text{time}$ among \{1, 2, 4, 8, 16, 32, 64, 128, 256\}. For baseline models, we use the best hyperparameters reported in ~\cite{LudewigJ18,LudewigMLJ19b}. Note that we conduct the reproducibility for previously reported results of all the baseline models. We implement the proposed model and STAN~\cite{GargGMVS19} using NumPy. We use public source code for SR-GNN\footnote{https://github.com/CRIPAC-DIG/SR-GNN} released by~\cite{WuT0WXT19} and the other baseline models\footnote{https://github.com/rn5l/session-rec} released by~\cite{LudewigMLJ19a}. (See the detailed hyperparameter settings in Appendix~\ref{sec:app_Reproducibility}.) We conduct all experiments on a desktop with 2 NVidia TITAN RTX, 256 GB memory, and 2 Intel Xeon Processor E5-2695 v4 (2.10 GHz, 45M cache). 

\section{Results and Discussion}\label{sec:results}

\label{sec:exp_time}

In this section, we discuss the accuracy and efficiency of our proposed models in comparison with eight baseline models. From the extensive experimental results, we summarize the following observations:

\begin{itemize}[leftmargin=5mm]
\item SLIST shows comparable or \textbf{state-of-the-art performance} on various datasets. For HR@20 and Recall@20, SLIST consistently outperforms baseline models. For MRR@20, SLIST is slightly worse than the best models on YC and RR datasets. Particularly, SLIST achieves up to 9.76\% and 8.83\% gain in HR@20 and MRR@20 over the best baseline model on the NowP dataset. (Section~\ref{sec:exp_overall})

\item Thanks to the closed-form solution, SLIST is up to \textbf{786x faster} in training than SR-GNN, a state-of-the-art DNN-based model. Also, the SLIST does not take significantly more time with larger training sets, indicating its superior \textbf{scalability}. (Section~\ref{sec:exp_time})

\item For relatively \textbf{shorter sessions}, SLIST significantly outperforms STAN and SR-GNN. While SLIT shows comparable performance with SLIST on the YC-1/4 dataset, it is much worse than SLIST on the DIGI1 dataset. (Section~\ref{sec:exp_length})

\item All the components of SLIST directly contribute to performance gains (\ie, up to 18.35\% improvement relative to whose components are all removed). The priority of components follows item weight decay ($\mathbf{w}_\text{inf}$) $>$ position decay ($\mathbf{w}_\text{pos}$) $>$ time decay ($\mathbf{w}_\text{time}$). (Section~\ref{sec:exp_ablation})
\end{itemize}

\subsection{Evaluation of Accuracy}
\label{sec:exp_overall}

Table~\ref{tab:overallresult} reports the accuracy of the proposed models and other competitors on seven datasets. (See the entire results for the other cut-offs (\ie, 5 and 10) in Appendix~\ref{sec:app_additionalresults}.)

First of all, we analyze the existing models based on the properties of datasets. We observe that no existing models ever establish state-of-the-art performance over all datasets. Either NARM or SR-GNN is the best model on YC datasets, but either SKNN or STAN is the best model on DIGI, RR, and NOWP datasets. This result implies that the performance of existing models suffers from high variance depending on the datasets. It is a natural consequence because different models mainly focus on a few different characteristics of sessions. DNN-based models commonly focus on sequential dependency, thus more advantageous to handle YC datasets presenting strong sequential dependency. On the other hand, neighborhood-based models generally show outstanding performance on DIGI5, RR, and NOWP datasets, showing stronger session consistency. This tendency explains that neighborhood-based models successfully exploit session consistency.

From this observation, we emphasize that achieving consistently higher performance across various datasets on session-based recommendation has been particularly challenging. Nevertheless, SLIST consistently records competitive performance, comparable to or better than state-of-the-art models. The consistent and competitive performance is a clear indicator that SLIST indeed effectively learns various aspects of sessions, not over-focusing on a subset of them. Specifically, SLIST achieves accuracy gains up to 9.76\% in HR@20 and 8.83 \% in MRR@20 over the best previous model. (We might achieve better performance for other metrics by tuning the hyperparameter $\alpha$ as shown in Figure~\ref{fig:combine_alpha}).

In most cases, SLIST outperforms its sub-components, SLIS and SLIT. It tells us that our unifying strategy is effective. Although some datasets might have a clear bias (\ie, one property is dominant while others are negligible), our joint training strategy still finds a reasonable solution considering the holistic aspects of the sessions. Similarly, STAN, an improved SKNN considering sequential dependency, also reports higher accuracy than SKNN. As a result, SLIST is the best or second-best model across all datasets.

Lastly, SEASE$^{R}$ does not show successful performance in the session-based scenario. Notably, SEASE$^{R}$ is much worse than SKNN, although both methods consider session consistency. It is because the diagonal penalty term in SEASE$^{R}$ prevents recommending the repeated items and leads to severe performance degradation. Therefore, we conclude that our modification to facilitate repeated item consumption is essential for building session-based recommendation models.

\subsection{Evaluation of Scalability}

\begin{table}[t] \small
\caption{Training time (in seconds) of SLIST and DNN-based models on the YC-1/4 dataset. Gains indicate how fast SLIST is compared to SR-GNN. While DNN-based models are trained using GPU, SLIST is trained using CPU.}\label{tab:time}

\begin{tabular}{c|rrrrr}
\toprule
\multirow{2}{*}{Models} & \multicolumn{5}{c}{The ratio of training sessions on YC-1/4}  \\
                        & 5\%     & 10\%     & 20\%     & 50\%     & 100\%     \\ 
\midrule
GRU4Rec$^{+}$           & 177.4  & 317.8   & 614.0   & 1600.2  & 3206.9    \\
NARM                    & 2137.6 & 3431.2  & 7454.2  & 27804.5 & 72076.8   \\
STAMP                   & 434.1  & 647.5   & 985.3   & 2081.2  & 4083.3    \\
SR-GNN                  & 6780.5 & 18014.7 & 31444.3 & 68581.9 & 185862.5 \\ \hline
SLIST                   & 202.7  & 199.0   & 199.9   & 227.0    & 241.9    \\ \hline
Gains                   & 33.5x   & 90.6x  & 157.3x  & 302.1x   & 768.2x    \\ 
\bottomrule
\end{tabular}
\end{table} 

Table~\ref{tab:time} compares the training time (in seconds) of SLIST and that of DNN-based models on the YC-1/4 dataset (\ie, the largest session dataset) with the various number of training sessions. Because the computational complexity of SLIST is mainly proportional to the number of items, the training time of SLIST is less dependent on the number of training sessions. For example, with 10\% of the entire training sessions, SLIST is only 90x faster than SR-GNN. Meanwhile, SLIST is 768x faster than SR-GNN with the entire training data. Also, SLIST takes a similar time to train on both the entire YC dataset and the YC-1/4 dataset. This is highly desirable in practice, especially on popular online services, where millions of users create billions of session data every day.

\subsection{Effect of Session Lengths}
\label{sec:exp_length}

\begin{figure}
\centering
\begin{tabular}{cc}
\includegraphics[width=0.220\textwidth]{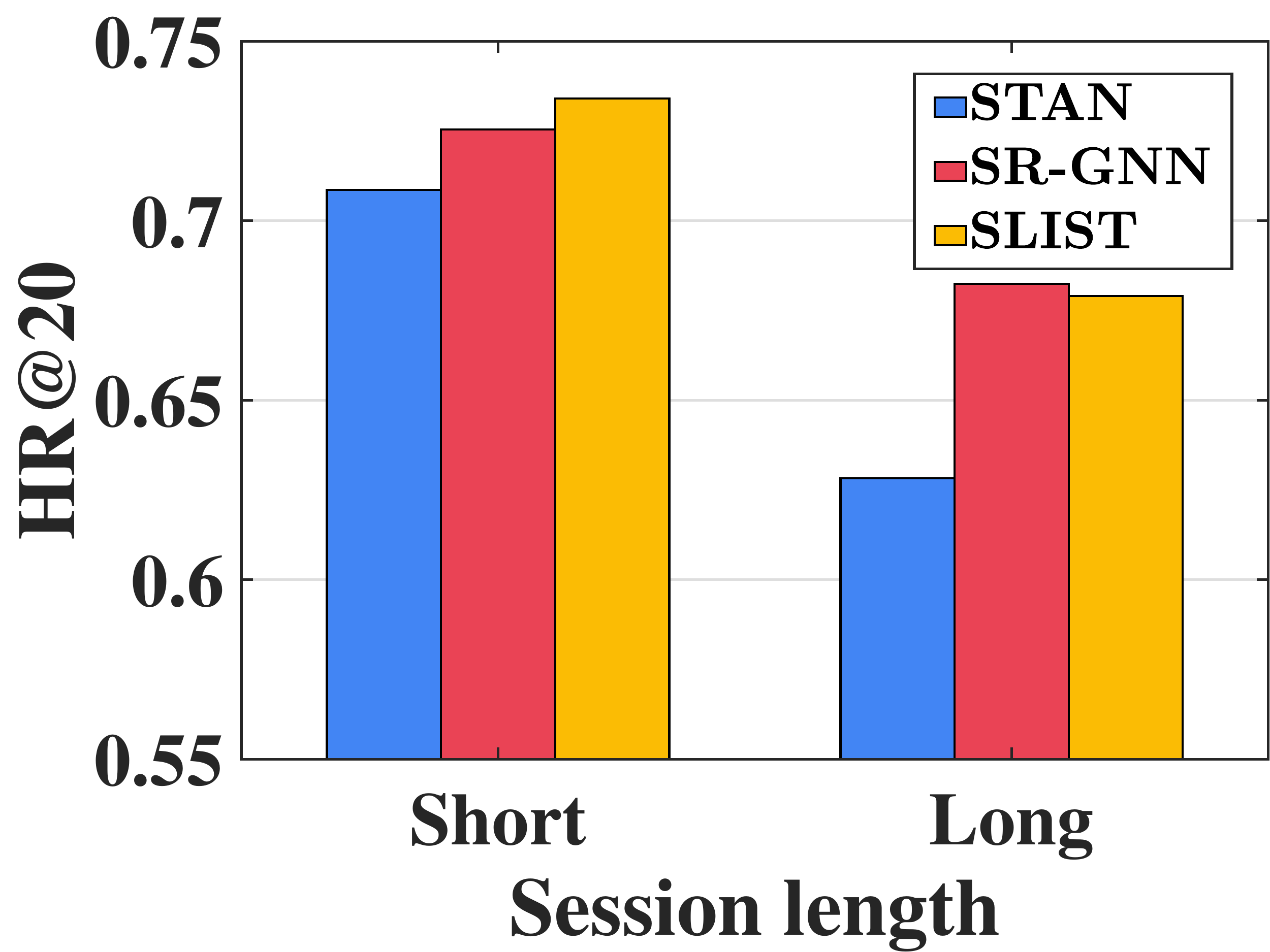} &
\includegraphics[width=0.220\textwidth]{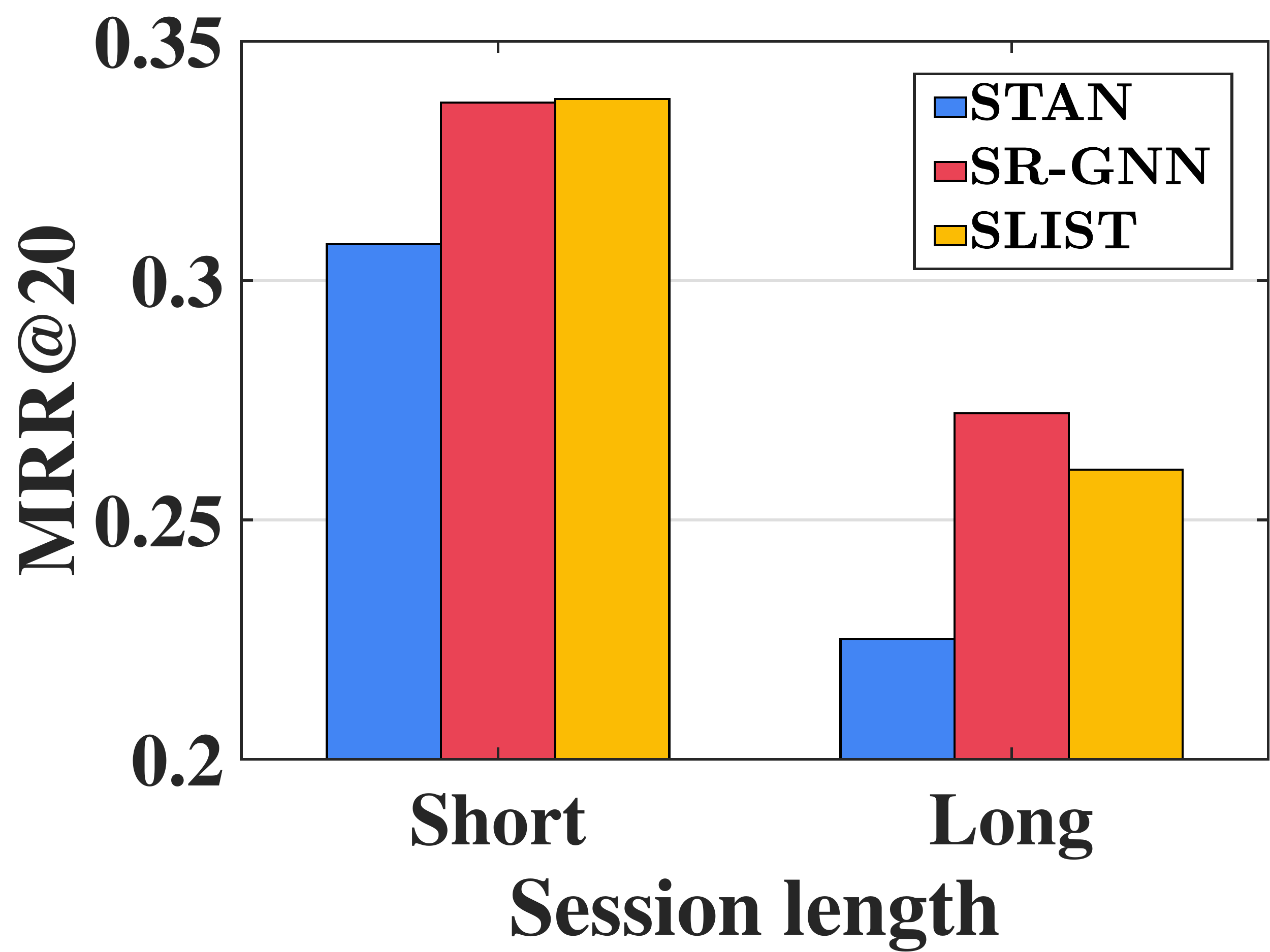} \\
\end{tabular}
(a) YC-1/4  \\
\begin{tabular}{cc}
\includegraphics[width=0.220\textwidth]{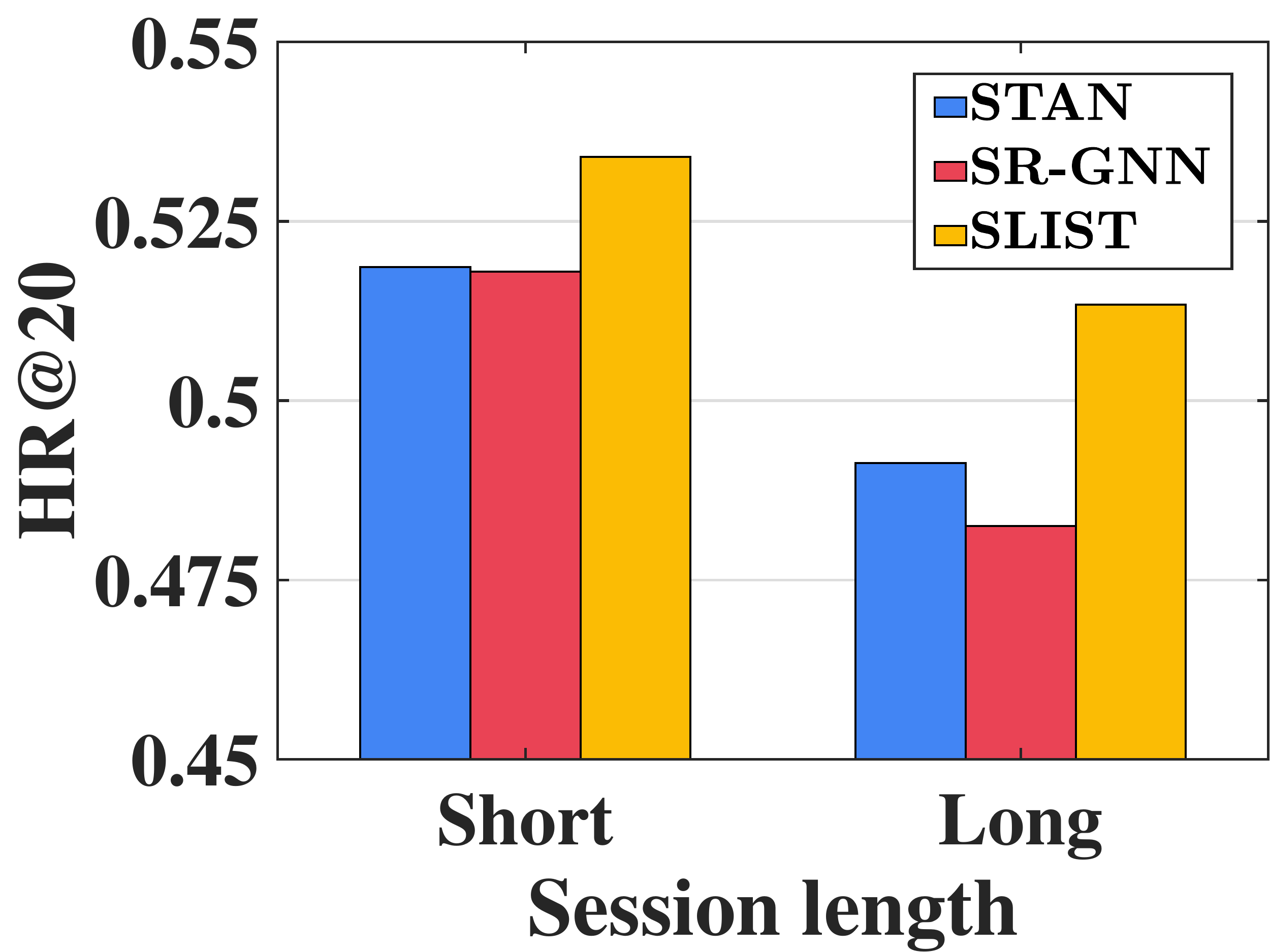} &
\includegraphics[width=0.220\textwidth]{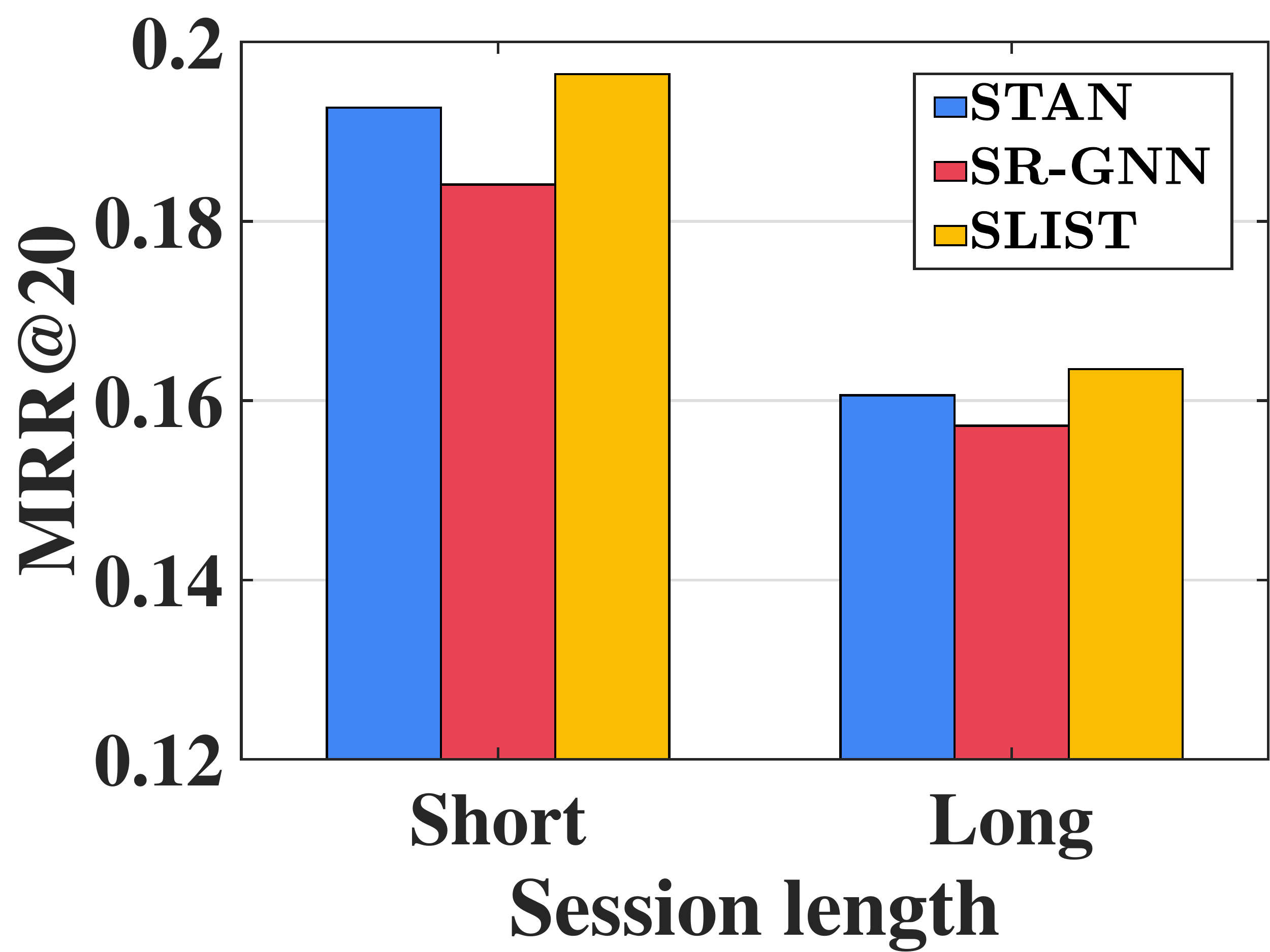} \\
\end{tabular}
(b) DIGI1  \\
\vskip -0.1in
\caption{HR@20 and MRR@20 of SLIST and state-of-the-art models with different session lengths (Short $\le$ 5, Long > 5) on YC-1/4 and DIGI1.}
\label{fig:length_baseline}
\end{figure}

Figure~\ref{fig:length_baseline} compares the accuracy of the proposed and state-of-the-art models (\ie, STAN and SR-GNN) under the short- (5 or fewer items) and long-session (more than 5 items) scenarios. The ratio of short sessions is 70.3\% (YC-1/4) and 76.4\% (DIGI1), respectively. First of all, all three methods show stronger performance in shorter sessions. This is probably because the user's intent may change over time, making it difficult to capture her intent. Second, the relative strength of each method tends to be larger with longer sessions, compared to STAN and SR-GNN. However, SLIST consistently outperforms both baselines in most scenarios, except for long sessions on YC-1/4, where SR-GNN slightly outperforms SLIST.

\begin{figure}
\centering
\begin{tabular}{cc}
\includegraphics[width=0.220\textwidth]{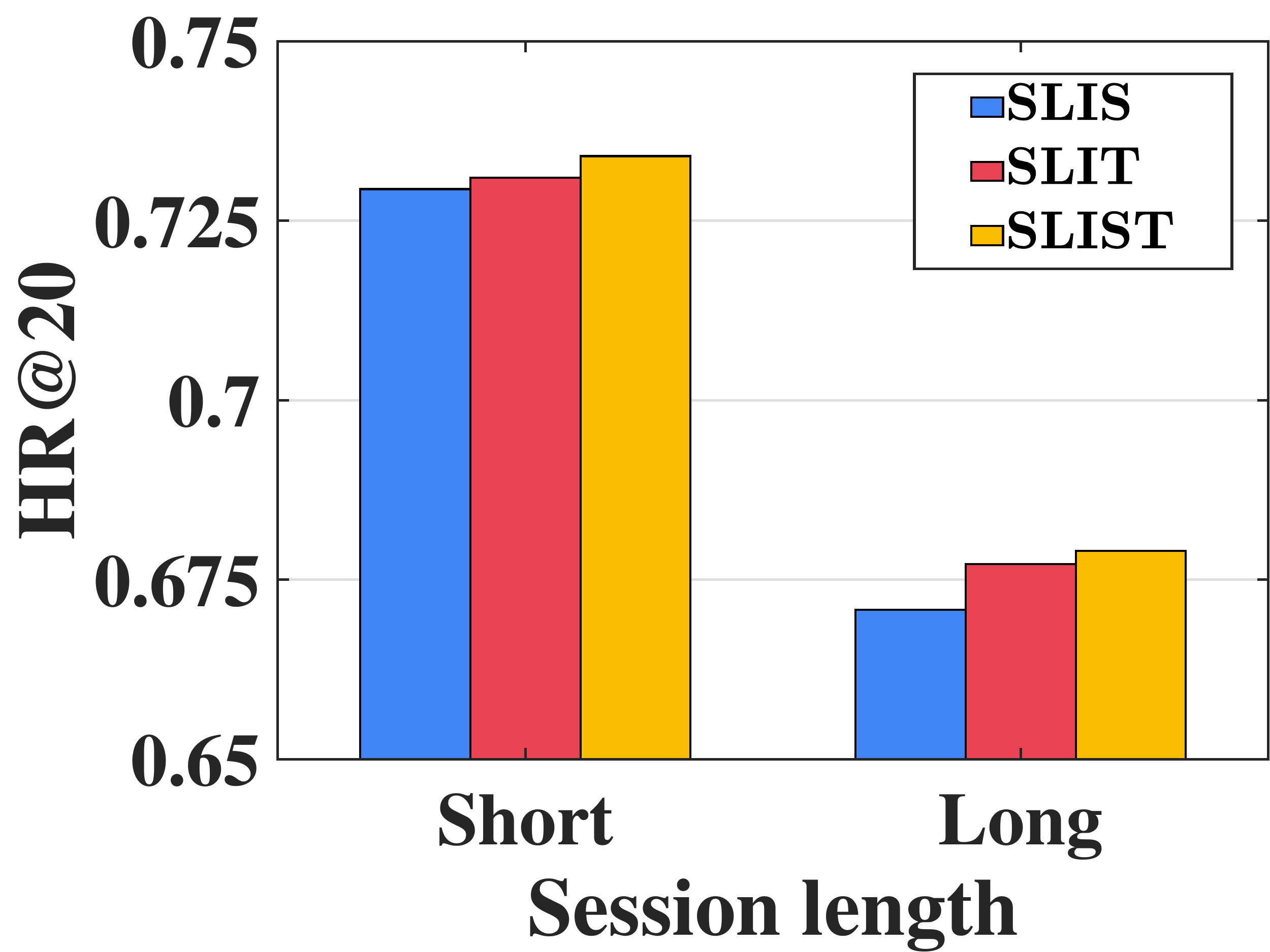} &
\includegraphics[width=0.220\textwidth]{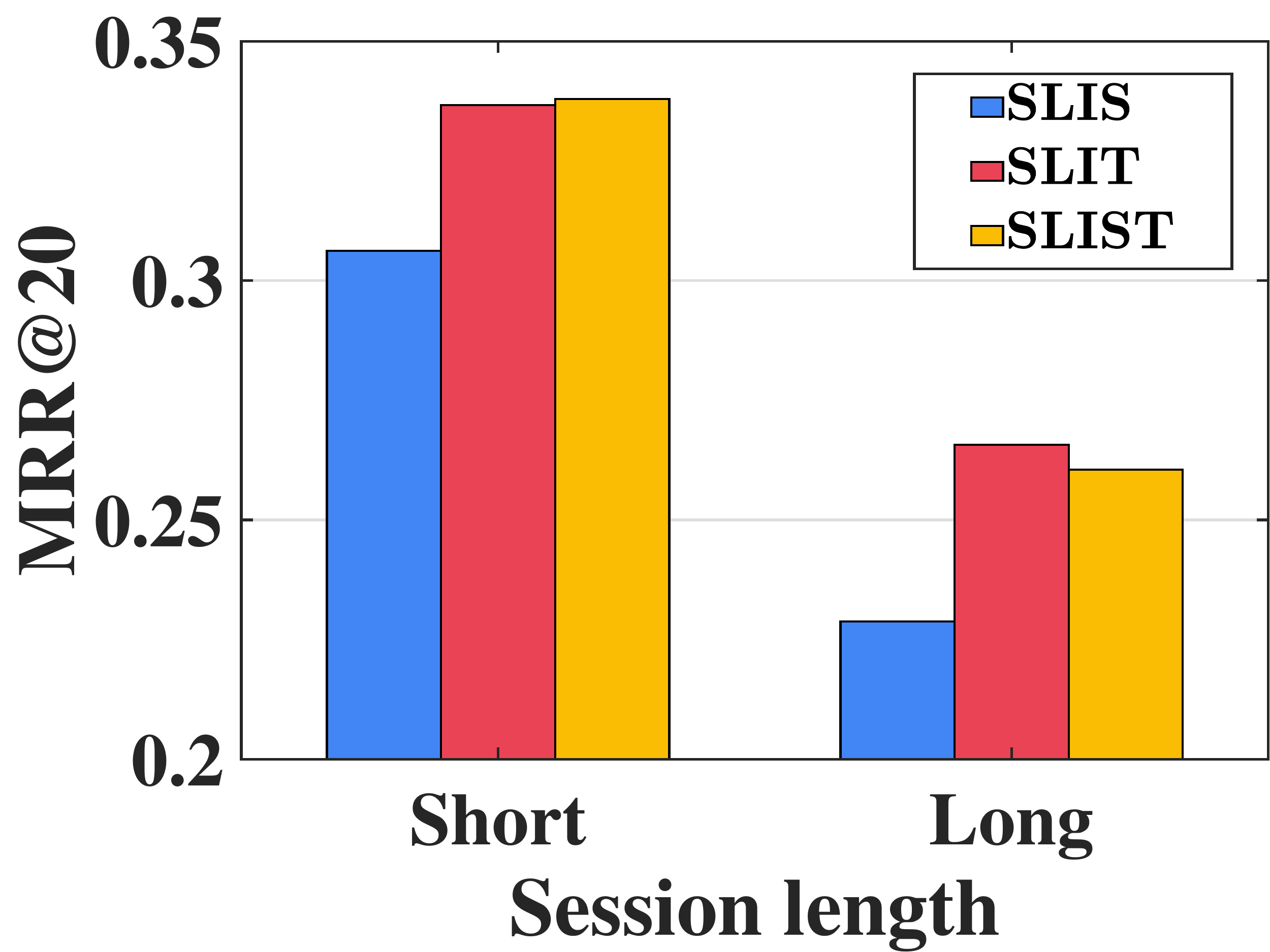} \\
\end{tabular}
(a) YC-1/4  \\
\begin{tabular}{cc}
\includegraphics[width=0.220\textwidth]{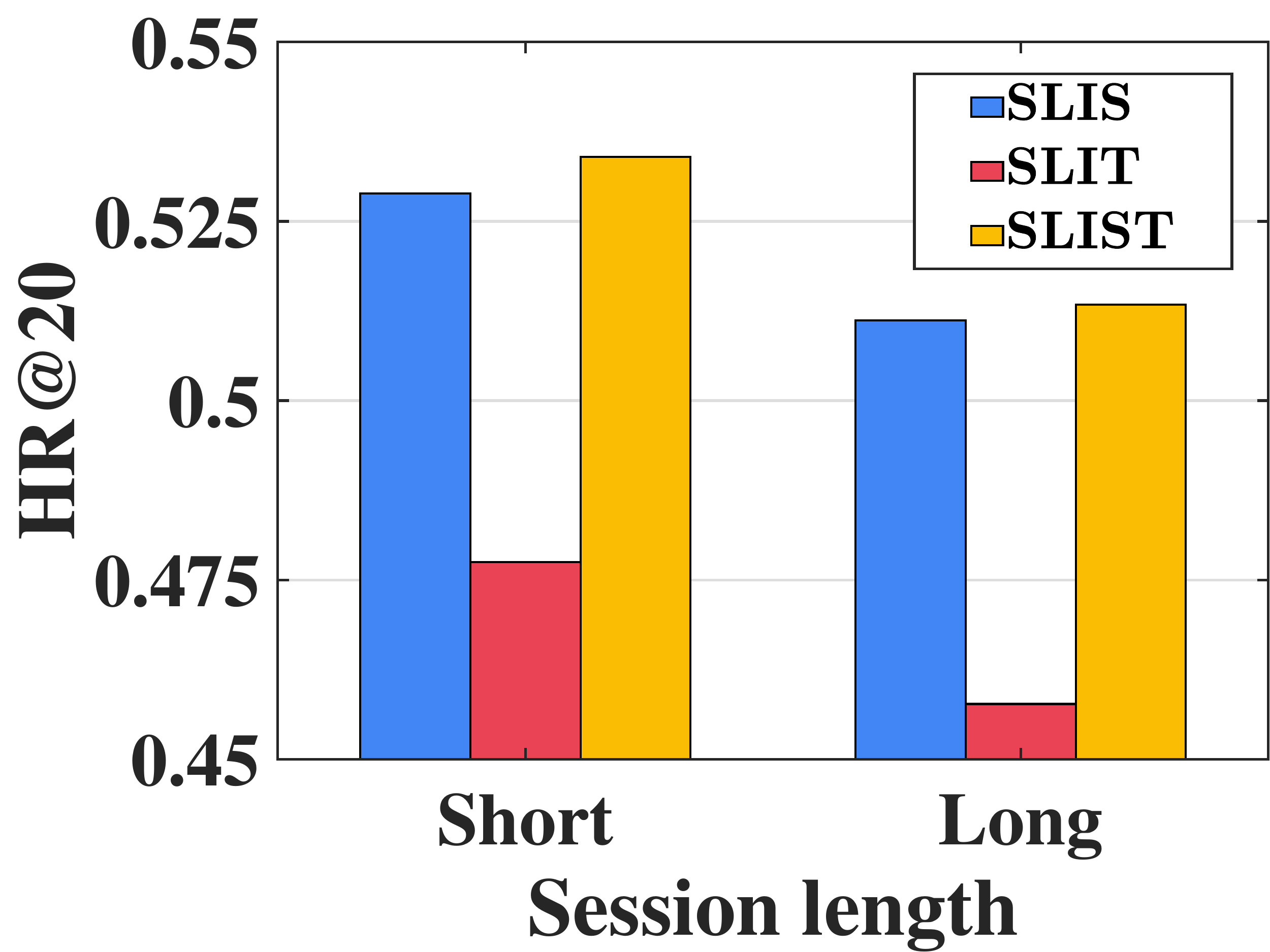} &
\includegraphics[width=0.220\textwidth]{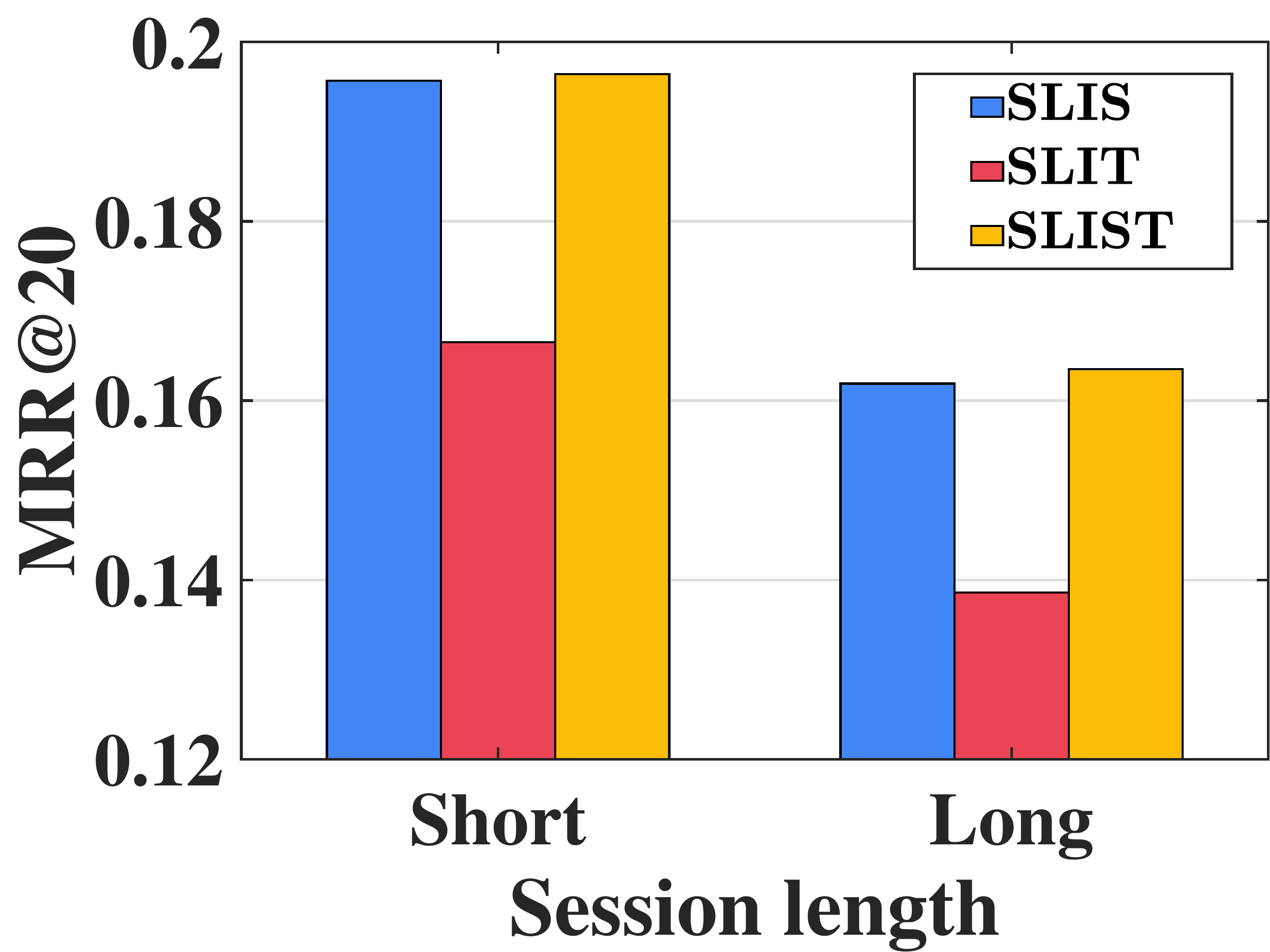} \\
\end{tabular}
(b) DIGI1  \\
\vskip -0.1in
\caption{HR@20 and MRR@20 of our three models with different session lengths (Short $\le$ 5, Long > 5) on YC-1/4 and DIGI1.}
\label{fig:length_proposed}
\vskip -0.15in
\end{figure}

Figure ~\ref{fig:length_proposed} compares the accuracy between our models. Compared to SLIS, SLIST improves the accuracy in MRR@20 up to 10.35\% and 0.35\% for the short sessions, and up to  13.85\% and 0.98\% for the long sessions, on YC-1/4 and DIGI1 datasets, respectively. Based on this experiment, we confirm that sequential dependency in YC-1/4 is prominent for long sessions because the user can change her intent during the session. This tendency is different from DIGI1 as SLIS outperforms SLIT for both short and long sessions. Consequently, we conclude that SLIST works well most flexibly on diverse datasets exhibiting highly different characteristics, while the baseline models and our sub-components (SLIS, SLIT) perform well only in a particular setting.

\subsection{Ablation Study}\label{sec:exp_ablation}

\begin{table}[t] \small
\caption{HR@20 and MRR@20 of SLIST when some components are removed: item weight decay $\mathbf{w}_\text{inf}$, position decay $\mathbf{w}_\text{pos}$, and time decay $\mathbf{w}_\text{time}$.}
\label{tab:ablation}
\begin{tabular}{ccc|cc|cc|cc}
\toprule
\multirow{2}{*}{\begin{turn}{90}$\mathbf{w}_\text{inf}$\end{turn}} & \multirow{2}{*}{\begin{turn}{90}$\mathbf{w}_\text{pos}$\end{turn}} & \multirow{2}{*}{\begin{turn}{90}$\mathbf{w}_\text{time}$\end{turn}} & \multicolumn{2}{c|}{YC-1/64} & \multicolumn{2}{c|}{YC-1/4} & \multicolumn{2}{c}{DIGI1} \\
    &    &    & HR            & MRR          & HR           & MRR          & HR           & MRR         \\
\midrule
\checkmark & \checkmark & \checkmark & \textbf{0.7088} & \textbf{0.3083} & \textbf{0.7175} & \textbf{0.3161} & \textbf{0.5291} & \textbf{0.1886}      \\
\checkmark & \checkmark &            & 0.7078  & 0.3077 & 0.7096 & 0.3115 & 0.5253 & 0.1880 \\
\checkmark &            & \checkmark & 0.6880  & 0.2973 & 0.7004 & 0.3031 & 0.5202 & 0.1855 \\
           & \checkmark & \checkmark & 0.6761  & 0.2697 & 0.6841 & 0.2753 & 0.5145 & 0.1820 \\
\checkmark &            &            & 0.6898  & 0.2980 & 0.6947 & 0.3005 & 0.5173 & 0.1852 \\
           & \checkmark &            & 0.6764 & 0.2697  & 0.6784 & 0.2723 & 0.5111 & 0.1817 \\
           &            & \checkmark & 0.6570  & 0.2651 & 0.6675 & 0.2693 & 0.5055 & 0.1789 \\
           &            &            & 0.6592  & 0.2652 & 0.6626 & 0.2671 & 0.5025 & 0.1786 \\
\bottomrule
\end{tabular}
\end{table}

We conducted an ablation study to analyze the effect of each component (\ie, position decay, item weight decay, time decay) in SLIST. For that, we removed each component from SLIST and observed performance degradation. Table~\ref{tab:ablation} summarizes the accuracy changes when each or a combination of components removed. 

First of all, having all components is always better than the models without one or more components. For all three datasets, the item weight decay is the most influential factor for accuracy. The second most influential one is position decay, and the least significant component is time decay. We consistently observe the same tendency over different datasets. This tendency can be interpreted as the factors for considering sequential dependency (\ie, item weight decay and position decay) is more crucial than the factor related to the timeliness of sessions (\ie, time decay). Also, time decay is no longer effective in YC-1/64 because it is the latest 1/16 out of YC-1/4, implying that YC-1/64 has already satisfied the timeliness of sessions.

Those three components are less significant in DIGI1, \eg, the gain from all three components in MRR@20 is 18.35\% and 5.59\% in YC-1/4 and DIGI1, respectively. As discussed earlier, the DIGI1 is more affected by session consistency than sequential dependency. For this reason, item weight decay and position decay are less influential in DIGI1 than those in YC datasets. (More experimental results and analysis on the effect of components are in Appendix ~\ref{sec:app_hyperparameters}.)

\begin{figure}[t]
\centering
\begin{tabular}{cc}
\includegraphics[width=0.23\textwidth]{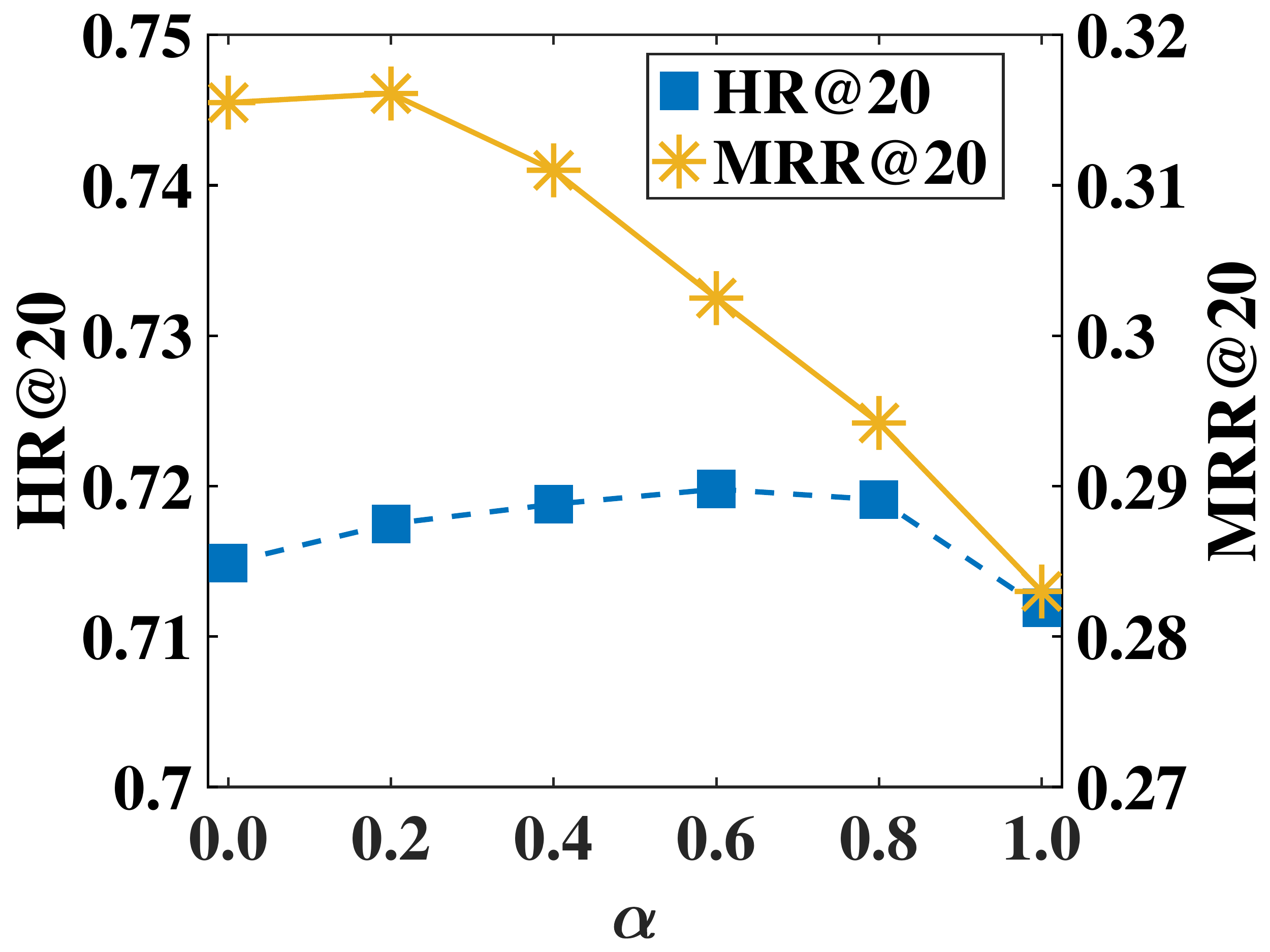} &
\includegraphics[width=0.23\textwidth]{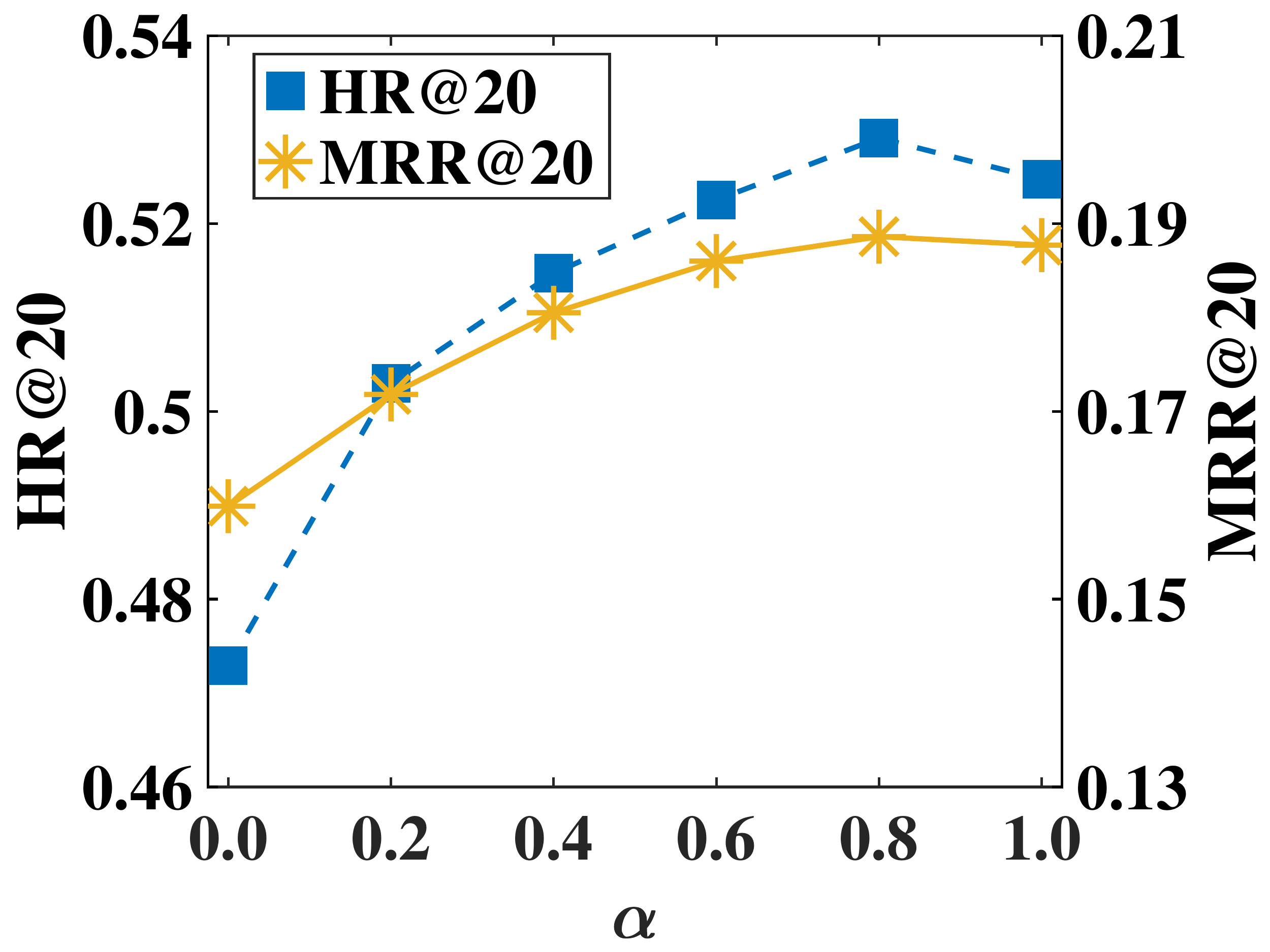} \\
(a) YC-1/4 & (b) DIGI1 \\
\end{tabular}
\vskip -0.1in
\caption{HR@20 and MRR@20 over varying $\alpha$ in SLIST on YC-1/4 and DIGI1.}
\label{fig:combine_alpha}
\vskip -0.15in
\end{figure}

Figure~\ref{fig:combine_alpha} depicts the effect of $\alpha$, controlling the importance of SLIS and SLIT in SLIST. SLIST with $\alpha=0$ is equivalent to SLIT, and SLIST with $\alpha=1$ is same as SLIS. For the datasets dominated by sequential dependency (\eg, YC-1/4), SLIT is more advantageous than SLIS. Thus, a small $\alpha$ is recommended, \eg, 0.2 is a good default value. On the contrary, when the target dataset exhibits session consistency more dominant than others (\eg, DIGI1), we observe better performance with higher $\alpha$ (\eg, 0.8), with more impact from SLIS.

The best performance is achieved with different $\alpha$, depending on the datasets and metrics. On YC-1/4, the best MRR@20 is achieved with $\alpha = 0.2$ (\ie, 0.3161), while the best HR@20 is with $\alpha = 0.8$ (\ie, 0.7198). As DIGI1 exhibits different characteristics of sessions, the best results are achieved when $\alpha = 0.8$ in both metrics.
\section{Related Work}\label{sec:relatedwork}

We briefly review session-based models with three categories. For more details, refer to recent survey papers~\cite{BonninJ14, JannachLL17}.

\vspace{1mm}
\noindent
\textbf{Markov-chain-based models}. Markov chains (MC) are widely used to model dependency from sequential data. FPMC~\cite{RendleFS10} proposed the tensor factorization that combines MC with traditional matrix factorization. Later, FOSSIL~\cite{HeM16} combined FISM~\cite{KabburNK13} with factorized MC. Recently, SR~\cite{KamehkhoshJL17} combined MC with association rules. Because MC-based models merely focus on the short-term dependency between items, they are limited for analyzing complex dependency between them.

\vspace{1mm}
\noindent
\textbf{DNN-based models}. Recently, session-based models have widely employed deep neural networks (DNNs), such as recurrent neural networks (RNNs), attention mechanisms, convolutional neural networks (CNNs), and graph neural networks (GNNs). As the pioneering work, ~\citet{HidasiKBT15} proposed GRU4Rec$^{+}$ with gated recurrent units (GRU) for session-based recommendation and then developed its optimized model~\cite{HidasiK18}. Later, NARM~\cite{LiRCRLM17} extended GRU4Rec$^{+}$ with an attention mechanism to capture short- and long-term dependency of items, and STAMP~\cite{LiuZMZ18} combined an attention mechanism with memory priority models to reflect the user's general interests. RepeatNet~\cite{RenCLR0R19} focused on analyzing the patterns for repeated items in the session. Recently, SR-GNN~\cite{WuT0WXT19} exploited gated graph neural networks (GGNN) to analyze the complex item transitions within a session. Then, GC-SAN~\cite{XuZLSXZFZ19} improved SR-GNN with a self-attention mechanism for contextualized non-local representations, and FGNN~\cite{PanCLR20a} adopted a weighted graph attention network (WGAT) to enhance SR-GNN. The GNN-based models effectively capture both the session consistency and sequential dependency in the session. Due to the extreme data sparsity, however, they are often vulnerable to overfitting~\cite{GuptaGMVS19}.

To summarize, DNN-based models show outstanding performance using the powerful model capacity. Because they usually incur slow training and inference time, however, it is difficult to support the scalability for large-scale datasets.

\vspace{1mm}
\noindent
\textbf{Neighborhood-based models}. To overcome the scalability issue, \citet{JannachL17} proposed SKNN that adopts the traditional K-nearest neighbor approach (KNN) for session recommendation, and STAN~\cite{GargGMVS19} improved SKNN using various weight schemes for sequential dependency and the recency of sessions. Later, ~\cite{LudewigJ18, LudewigMLJ19a, LudewigMLJ19b} extensively conducted an empirical study with various session-based models. Surprisingly, although non-neural models are simple, they show competitive performance in several benchmark datasets. To leverage the correlations between inter-session items, recent studies (\eg, CSRM~\cite{WangRMCMR19} and CoSAN~\cite{LuoZLZWXFS20}) incorporated neighbor sessions into DNN-based models. While KNN-based models can capture session consistency, they are generally limited to represent complex dependency. Besides, they are also sensitive to similarity metrics between sessions.

\section{Conclusion}\label{sec:conclusion}

This paper presents \emph{Session-aware Linear Item Similarity/Transition model (SLIST)} to tackle session-based recommendation. To complement the drawback of existing models, we unify two linear models with different perspectives to fully capture various characteristics of sessions. To the best of our knowledge, our model is the first work that adopts the linear item-item models to fully utilize various characteristics of sessions. Owing to its closed-form solution, SLIST is also highly scalable. Through comprehensive evaluation, we demonstrate that SLIST achieves comparable or state-of-the-art accuracy over existing DNN-based and neighborhood-based models on multiple benchmark datasets.

\section*{Acknowledgment}

This work was supported by the National Research Foundation of Korea (NRF) (NRF-2018R1A5A1060031). Also, this work was supported by Institute of Information \& communications Technology Planning \& evaluation (IITP) grant funded by the Korea government (MSIT) (No.2019-0-00421, AI Graduate School Support Program and IITP-2020-0-01821, ICT Creative Consilience Program).

\bibliographystyle{ACM-Reference-Format}
\bibliography{references}


\begin{thebibliography}{45}


\ifx \showCODEN    \undefined \def \showCODEN     #1{\unskip}     \fi
\ifx \showDOI      \undefined \def \showDOI       #1{#1}\fi
\ifx \showISBNx    \undefined \def \showISBNx     #1{\unskip}     \fi
\ifx \showISBNxiii \undefined \def \showISBNxiii  #1{\unskip}     \fi
\ifx \showISSN     \undefined \def \showISSN      #1{\unskip}     \fi
\ifx \showLCCN     \undefined \def \showLCCN      #1{\unskip}     \fi
\ifx \shownote     \undefined \def \shownote      #1{#1}          \fi
\ifx \showarticletitle \undefined \def \showarticletitle #1{#1}   \fi
\ifx \showURL      \undefined \def \showURL       {\relax}        \fi
\providecommand\bibfield[2]{#2}
\providecommand\bibinfo[2]{#2}
\providecommand\natexlab[1]{#1}
\providecommand\showeprint[2][]{arXiv:#2}

\bibitem[\protect\citeauthoryear{Bonnin and Jannach}{Bonnin and
  Jannach}{2014}]%
        {BonninJ14}
\bibfield{author}{\bibinfo{person}{Geoffray Bonnin} {and}
  \bibinfo{person}{Dietmar Jannach}.} \bibinfo{year}{2014}\natexlab{}.
\newblock \showarticletitle{Automated Generation of Music Playlists: Survey and
  Experiments}.
\newblock \bibinfo{journal}{\emph{{ACM} Comput. Surv.}} \bibinfo{volume}{47},
  \bibinfo{number}{2} (\bibinfo{year}{2014}), \bibinfo{pages}{26:1--26:35}.
\newblock


\bibitem[\protect\citeauthoryear{Choi, Jeong, Lee, and Lee}{Choi
  et~al\mbox{.}}{2021}]%
        {ChoiJLL21}
\bibfield{author}{\bibinfo{person}{Minjin Choi}, \bibinfo{person}{Yoonki
  Jeong}, \bibinfo{person}{Joonseok Lee}, {and} \bibinfo{person}{Jongwuk Lee}.}
  \bibinfo{year}{2021}\natexlab{}.
\newblock \showarticletitle{Local Collaborative Autoencoders}. In
  \bibinfo{booktitle}{\emph{WSDM}}.
\newblock


\bibitem[\protect\citeauthoryear{Fang, Guo, Zhang, and Shu}{Fang
  et~al\mbox{.}}{2019}]%
        {FangGZS19}
\bibfield{author}{\bibinfo{person}{Hui Fang}, \bibinfo{person}{Guibing Guo},
  \bibinfo{person}{Danning Zhang}, {and} \bibinfo{person}{Yiheng Shu}.}
  \bibinfo{year}{2019}\natexlab{}.
\newblock \showarticletitle{Deep Learning-Based Sequential Recommender Systems:
  Concepts, Algorithms, and Evaluations}. In \bibinfo{booktitle}{\emph{ICWE}}.
  \bibinfo{pages}{574--577}.
\newblock


\bibitem[\protect\citeauthoryear{Garg, Gupta, Malhotra, Vig, and Shroff}{Garg
  et~al\mbox{.}}{2019}]%
        {GargGMVS19}
\bibfield{author}{\bibinfo{person}{Diksha Garg}, \bibinfo{person}{Priyanka
  Gupta}, \bibinfo{person}{Pankaj Malhotra}, \bibinfo{person}{Lovekesh Vig},
  {and} \bibinfo{person}{Gautam~M. Shroff}.} \bibinfo{year}{2019}\natexlab{}.
\newblock \showarticletitle{Sequence and Time Aware Neighborhood for
  Session-based Recommendations: {STAN}}. In \bibinfo{booktitle}{\emph{SIGIR}}.
  \bibinfo{pages}{1069--1072}.
\newblock


\bibitem[\protect\citeauthoryear{Goldberg, Nichols, Oki, and Terry}{Goldberg
  et~al\mbox{.}}{1992}]%
        {GoldbergNOT92}
\bibfield{author}{\bibinfo{person}{David Goldberg}, \bibinfo{person}{David~A.
  Nichols}, \bibinfo{person}{Brian~M. Oki}, {and} \bibinfo{person}{Douglas~B.
  Terry}.} \bibinfo{year}{1992}\natexlab{}.
\newblock \showarticletitle{Using Collaborative Filtering to Weave an
  Information Tapestry}.
\newblock \bibinfo{journal}{\emph{Commun. {ACM}}} \bibinfo{volume}{35},
  \bibinfo{number}{12} (\bibinfo{year}{1992}), \bibinfo{pages}{61--70}.
\newblock


\bibitem[\protect\citeauthoryear{Gupta, Garg, Malhotra, Vig, and Shroff}{Gupta
  et~al\mbox{.}}{2019}]%
        {GuptaGMVS19}
\bibfield{author}{\bibinfo{person}{Priyanka Gupta}, \bibinfo{person}{Diksha
  Garg}, \bibinfo{person}{Pankaj Malhotra}, \bibinfo{person}{Lovekesh Vig},
  {and} \bibinfo{person}{Gautam~M. Shroff}.} \bibinfo{year}{2019}\natexlab{}.
\newblock \showarticletitle{{NISER:} Normalized Item and Session
  Representations with Graph Neural Networks}.
\newblock \bibinfo{journal}{\emph{CoRR}}  \bibinfo{volume}{abs/1909.04276}
  (\bibinfo{year}{2019}).
\newblock


\bibitem[\protect\citeauthoryear{He and McAuley}{He and McAuley}{2016}]%
        {HeM16}
\bibfield{author}{\bibinfo{person}{Ruining He} {and} \bibinfo{person}{Julian~J.
  McAuley}.} \bibinfo{year}{2016}\natexlab{}.
\newblock \showarticletitle{Fusing Similarity Models with Markov Chains for
  Sparse Sequential Recommendation}. In \bibinfo{booktitle}{\emph{ICDM}}.
  \bibinfo{pages}{191--200}.
\newblock


\bibitem[\protect\citeauthoryear{He, Liao, Zhang, Nie, Hu, and Chua}{He
  et~al\mbox{.}}{2017}]%
        {HeLZNHC17}
\bibfield{author}{\bibinfo{person}{Xiangnan He}, \bibinfo{person}{Lizi Liao},
  \bibinfo{person}{Hanwang Zhang}, \bibinfo{person}{Liqiang Nie},
  \bibinfo{person}{Xia Hu}, {and} \bibinfo{person}{Tat{-}Seng Chua}.}
  \bibinfo{year}{2017}\natexlab{}.
\newblock \showarticletitle{Neural Collaborative Filtering}. In
  \bibinfo{booktitle}{\emph{WWW}}. \bibinfo{pages}{173--182}.
\newblock


\bibitem[\protect\citeauthoryear{Herlocker, Konstan, Borchers, and
  Riedl}{Herlocker et~al\mbox{.}}{1999}]%
        {HerlockerKBR99}
\bibfield{author}{\bibinfo{person}{Jonathan~L. Herlocker},
  \bibinfo{person}{Joseph~A. Konstan}, \bibinfo{person}{Al Borchers}, {and}
  \bibinfo{person}{John Riedl}.} \bibinfo{year}{1999}\natexlab{}.
\newblock \showarticletitle{An Algorithmic Framework for Performing
  Collaborative Filtering}. In \bibinfo{booktitle}{\emph{SIGIR}}.
  \bibinfo{pages}{230--237}.
\newblock


\bibitem[\protect\citeauthoryear{Hidasi and Karatzoglou}{Hidasi and
  Karatzoglou}{2018}]%
        {HidasiK18}
\bibfield{author}{\bibinfo{person}{Bal{\'{a}}zs Hidasi} {and}
  \bibinfo{person}{Alexandros Karatzoglou}.} \bibinfo{year}{2018}\natexlab{}.
\newblock \showarticletitle{Recurrent Neural Networks with Top-k Gains for
  Session-based Recommendations}. In \bibinfo{booktitle}{\emph{CIKM}}.
  \bibinfo{pages}{843--852}.
\newblock


\bibitem[\protect\citeauthoryear{Hidasi, Karatzoglou, Baltrunas, and
  Tikk}{Hidasi et~al\mbox{.}}{2016a}]%
        {HidasiKBT15}
\bibfield{author}{\bibinfo{person}{Bal{\'{a}}zs Hidasi},
  \bibinfo{person}{Alexandros Karatzoglou}, \bibinfo{person}{Linas Baltrunas},
  {and} \bibinfo{person}{Domonkos Tikk}.} \bibinfo{year}{2016}\natexlab{a}.
\newblock \showarticletitle{Session-based Recommendations with Recurrent Neural
  Networks}. In \bibinfo{booktitle}{\emph{ICLR}}.
\newblock


\bibitem[\protect\citeauthoryear{Hidasi, Quadrana, Karatzoglou, and
  Tikk}{Hidasi et~al\mbox{.}}{2016b}]%
        {HidasiQKT16}
\bibfield{author}{\bibinfo{person}{Bal{\'{a}}zs Hidasi},
  \bibinfo{person}{Massimo Quadrana}, \bibinfo{person}{Alexandros Karatzoglou},
  {and} \bibinfo{person}{Domonkos Tikk}.} \bibinfo{year}{2016}\natexlab{b}.
\newblock \showarticletitle{Parallel Recurrent Neural Network Architectures for
  Feature-rich Session-based Recommendations}. In
  \bibinfo{booktitle}{\emph{RecSys}}. \bibinfo{pages}{241--248}.
\newblock


\bibitem[\protect\citeauthoryear{Hu, Koren, and Volinsky}{Hu
  et~al\mbox{.}}{2008}]%
        {HuKV08}
\bibfield{author}{\bibinfo{person}{Yifan Hu}, \bibinfo{person}{Yehuda Koren},
  {and} \bibinfo{person}{Chris Volinsky}.} \bibinfo{year}{2008}\natexlab{}.
\newblock \showarticletitle{Collaborative Filtering for Implicit Feedback
  Datasets}. In \bibinfo{booktitle}{\emph{ICDM}}. \bibinfo{pages}{263--272}.
\newblock


\bibitem[\protect\citeauthoryear{Jannach and Ludewig}{Jannach and
  Ludewig}{2017}]%
        {JannachL17}
\bibfield{author}{\bibinfo{person}{Dietmar Jannach} {and}
  \bibinfo{person}{Malte Ludewig}.} \bibinfo{year}{2017}\natexlab{}.
\newblock \showarticletitle{When Recurrent Neural Networks meet the
  Neighborhood for Session-Based Recommendation}. In
  \bibinfo{booktitle}{\emph{RecSys}}. \bibinfo{pages}{306--310}.
\newblock


\bibitem[\protect\citeauthoryear{Jannach, Ludewig, and Lerche}{Jannach
  et~al\mbox{.}}{2017}]%
        {JannachLL17}
\bibfield{author}{\bibinfo{person}{Dietmar Jannach}, \bibinfo{person}{Malte
  Ludewig}, {and} \bibinfo{person}{Lukas Lerche}.}
  \bibinfo{year}{2017}\natexlab{}.
\newblock \showarticletitle{Session-based item recommendation in e-commerce: on
  short-term intents, reminders, trends and discounts}.
\newblock \bibinfo{journal}{\emph{User Model. User Adapt. Interact.}}
  \bibinfo{volume}{27}, \bibinfo{number}{3-5} (\bibinfo{year}{2017}),
  \bibinfo{pages}{351--392}.
\newblock


\bibitem[\protect\citeauthoryear{Jeunen, Balen, and Goethals}{Jeunen
  et~al\mbox{.}}{2020}]%
        {JeunenBG20}
\bibfield{author}{\bibinfo{person}{Olivier Jeunen}, \bibinfo{person}{Jan~Van
  Balen}, {and} \bibinfo{person}{Bart Goethals}.}
  \bibinfo{year}{2020}\natexlab{}.
\newblock \showarticletitle{Closed-Form Models for Collaborative Filtering with
  Side-Information}. In \bibinfo{booktitle}{\emph{RecSys}}.
  \bibinfo{pages}{651--656}.
\newblock


\bibitem[\protect\citeauthoryear{Kabbur, Ning, and Karypis}{Kabbur
  et~al\mbox{.}}{2013}]%
        {KabburNK13}
\bibfield{author}{\bibinfo{person}{Santosh Kabbur}, \bibinfo{person}{Xia Ning},
  {and} \bibinfo{person}{George Karypis}.} \bibinfo{year}{2013}\natexlab{}.
\newblock \showarticletitle{{FISM:} factored item similarity models for top-N
  recommender systems}. In \bibinfo{booktitle}{\emph{KDD}}.
  \bibinfo{pages}{659--667}.
\newblock


\bibitem[\protect\citeauthoryear{Kamehkhosh, Jannach, and Ludewig}{Kamehkhosh
  et~al\mbox{.}}{2017}]%
        {KamehkhoshJL17}
\bibfield{author}{\bibinfo{person}{Iman Kamehkhosh}, \bibinfo{person}{Dietmar
  Jannach}, {and} \bibinfo{person}{Malte Ludewig}.}
  \bibinfo{year}{2017}\natexlab{}.
\newblock \showarticletitle{A Comparison of Frequent Pattern Techniques and a
  Deep Learning Method for Session-Based Recommendation}. In
  \bibinfo{booktitle}{\emph{RecSys}}. \bibinfo{pages}{50--56}.
\newblock


\bibitem[\protect\citeauthoryear{Lee, Abu{-}El{-}Haija, Varadarajan, and
  Natsev}{Lee et~al\mbox{.}}{2018}]%
        {LeeAVN18}
\bibfield{author}{\bibinfo{person}{Joonseok Lee}, \bibinfo{person}{Sami
  Abu{-}El{-}Haija}, \bibinfo{person}{Balakrishnan Varadarajan}, {and}
  \bibinfo{person}{Apostol Natsev}.} \bibinfo{year}{2018}\natexlab{}.
\newblock \showarticletitle{Collaborative Deep Metric Learning for Video
  Understanding}. \bibinfo{pages}{481--490}.
\newblock


\bibitem[\protect\citeauthoryear{Lee, Bengio, Kim, Lebanon, and Singer}{Lee
  et~al\mbox{.}}{2014}]%
        {LeeBKLS14}
\bibfield{author}{\bibinfo{person}{Joonseok Lee}, \bibinfo{person}{Samy
  Bengio}, \bibinfo{person}{Seungyeon Kim}, \bibinfo{person}{Guy Lebanon},
  {and} \bibinfo{person}{Yoram Singer}.} \bibinfo{year}{2014}\natexlab{}.
\newblock \showarticletitle{Local collaborative ranking}.
  \bibinfo{pages}{85--96}.
\newblock


\bibitem[\protect\citeauthoryear{Lee, Kim, Lebanon, Singer, and Bengio}{Lee
  et~al\mbox{.}}{2016}]%
        {LeeKLSB16}
\bibfield{author}{\bibinfo{person}{Joonseok Lee}, \bibinfo{person}{Seungyeon
  Kim}, \bibinfo{person}{Guy Lebanon}, \bibinfo{person}{Yoram Singer}, {and}
  \bibinfo{person}{Samy Bengio}.} \bibinfo{year}{2016}\natexlab{}.
\newblock \showarticletitle{{LLORMA:} Local Low-Rank Matrix Approximation}.
\newblock \bibinfo{journal}{\emph{Journal of Machine Learning Research}}
  \bibinfo{volume}{17} (\bibinfo{year}{2016}), \bibinfo{pages}{15:1--15:24}.
\newblock


\bibitem[\protect\citeauthoryear{Li, Ren, Chen, Ren, Lian, and Ma}{Li
  et~al\mbox{.}}{2017}]%
        {LiRCRLM17}
\bibfield{author}{\bibinfo{person}{Jing Li}, \bibinfo{person}{Pengjie Ren},
  \bibinfo{person}{Zhumin Chen}, \bibinfo{person}{Zhaochun Ren},
  \bibinfo{person}{Tao Lian}, {and} \bibinfo{person}{Jun Ma}.}
  \bibinfo{year}{2017}\natexlab{}.
\newblock \showarticletitle{Neural Attentive Session-based Recommendation}. In
  \bibinfo{booktitle}{\emph{CIKM}}. \bibinfo{pages}{1419--1428}.
\newblock


\bibitem[\protect\citeauthoryear{Liang, Krishnan, Hoffman, and Jebara}{Liang
  et~al\mbox{.}}{2018}]%
        {LiangKHJ18}
\bibfield{author}{\bibinfo{person}{Dawen Liang}, \bibinfo{person}{Rahul~G.
  Krishnan}, \bibinfo{person}{Matthew~D. Hoffman}, {and} \bibinfo{person}{Tony
  Jebara}.} \bibinfo{year}{2018}\natexlab{}.
\newblock \showarticletitle{Variational Autoencoders for Collaborative
  Filtering}. In \bibinfo{booktitle}{\emph{WWW}}. \bibinfo{pages}{689--698}.
\newblock


\bibitem[\protect\citeauthoryear{Liu, Zeng, Mokhosi, and Zhang}{Liu
  et~al\mbox{.}}{2018}]%
        {LiuZMZ18}
\bibfield{author}{\bibinfo{person}{Qiao Liu}, \bibinfo{person}{Yifu Zeng},
  \bibinfo{person}{Refuoe Mokhosi}, {and} \bibinfo{person}{Haibin Zhang}.}
  \bibinfo{year}{2018}\natexlab{}.
\newblock \showarticletitle{{STAMP:} Short-Term Attention/Memory Priority Model
  for Session-based Recommendation}. In \bibinfo{booktitle}{\emph{KDD}}.
  \bibinfo{pages}{1831--1839}.
\newblock


\bibitem[\protect\citeauthoryear{Ludewig and Jannach}{Ludewig and
  Jannach}{2018}]%
        {LudewigJ18}
\bibfield{author}{\bibinfo{person}{Malte Ludewig} {and}
  \bibinfo{person}{Dietmar Jannach}.} \bibinfo{year}{2018}\natexlab{}.
\newblock \showarticletitle{Evaluation of session-based recommendation
  algorithms}.
\newblock \bibinfo{journal}{\emph{User Model. User Adapt. Interact.}}
  \bibinfo{volume}{28}, \bibinfo{number}{4-5} (\bibinfo{year}{2018}),
  \bibinfo{pages}{331--390}.
\newblock


\bibitem[\protect\citeauthoryear{Ludewig, Mauro, Latifi, and Jannach}{Ludewig
  et~al\mbox{.}}{2019a}]%
        {LudewigMLJ19b}
\bibfield{author}{\bibinfo{person}{Malte Ludewig}, \bibinfo{person}{Noemi
  Mauro}, \bibinfo{person}{Sara Latifi}, {and} \bibinfo{person}{Dietmar
  Jannach}.} \bibinfo{year}{2019}\natexlab{a}.
\newblock \showarticletitle{Empirical Analysis of Session-Based Recommendation
  Algorithms}.
\newblock \bibinfo{journal}{\emph{CoRR}}  \bibinfo{volume}{abs/1910.12781}
  (\bibinfo{year}{2019}).
\newblock


\bibitem[\protect\citeauthoryear{Ludewig, Mauro, Latifi, and Jannach}{Ludewig
  et~al\mbox{.}}{2019b}]%
        {LudewigMLJ19a}
\bibfield{author}{\bibinfo{person}{Malte Ludewig}, \bibinfo{person}{Noemi
  Mauro}, \bibinfo{person}{Sara Latifi}, {and} \bibinfo{person}{Dietmar
  Jannach}.} \bibinfo{year}{2019}\natexlab{b}.
\newblock \showarticletitle{Performance comparison of neural and non-neural
  approaches to session-based recommendation}. In
  \bibinfo{booktitle}{\emph{RecSys}}. \bibinfo{pages}{462--466}.
\newblock


\bibitem[\protect\citeauthoryear{Luo, Zhao, Liu, Zhuang, Wang, Xu, Fang, and
  Sheng}{Luo et~al\mbox{.}}{2020}]%
        {LuoZLZWXFS20}
\bibfield{author}{\bibinfo{person}{Anjing Luo}, \bibinfo{person}{Pengpeng
  Zhao}, \bibinfo{person}{Yanchi Liu}, \bibinfo{person}{Fuzhen Zhuang},
  \bibinfo{person}{Deqing Wang}, \bibinfo{person}{Jiajie Xu},
  \bibinfo{person}{Junhua Fang}, {and} \bibinfo{person}{Victor~S. Sheng}.}
  \bibinfo{year}{2020}\natexlab{}.
\newblock \showarticletitle{Collaborative Self-Attention Network for
  Session-based Recommendation}. In \bibinfo{booktitle}{\emph{IJCAI}}.
  \bibinfo{pages}{2591--2597}.
\newblock


\bibitem[\protect\citeauthoryear{Ning and Karypis}{Ning and Karypis}{2011}]%
        {NingK11}
\bibfield{author}{\bibinfo{person}{Xia Ning} {and} \bibinfo{person}{George
  Karypis}.} \bibinfo{year}{2011}\natexlab{}.
\newblock \showarticletitle{{SLIM:} Sparse Linear Methods for Top-N Recommender
  Systems}. In \bibinfo{booktitle}{\emph{ICDM}}. \bibinfo{pages}{497--506}.
\newblock


\bibitem[\protect\citeauthoryear{Pan, Cai, Ling, and de~Rijke}{Pan
  et~al\mbox{.}}{2020}]%
        {PanCLR20a}
\bibfield{author}{\bibinfo{person}{Zhiqiang Pan}, \bibinfo{person}{Fei Cai},
  \bibinfo{person}{Yanxiang Ling}, {and} \bibinfo{person}{Maarten de Rijke}.}
  \bibinfo{year}{2020}\natexlab{}.
\newblock \showarticletitle{Rethinking Item Importance in Session-based
  Recommendation}. In \bibinfo{booktitle}{\emph{SIGIR}}.
  \bibinfo{pages}{1837--1840}.
\newblock


\bibitem[\protect\citeauthoryear{Quadrana, Cremonesi, and Jannach}{Quadrana
  et~al\mbox{.}}{2018}]%
        {QuadranaCJ18}
\bibfield{author}{\bibinfo{person}{Massimo Quadrana}, \bibinfo{person}{Paolo
  Cremonesi}, {and} \bibinfo{person}{Dietmar Jannach}.}
  \bibinfo{year}{2018}\natexlab{}.
\newblock \showarticletitle{Sequence-Aware Recommender Systems}.
\newblock \bibinfo{journal}{\emph{{ACM} Comput. Surv.}} \bibinfo{volume}{51},
  \bibinfo{number}{4} (\bibinfo{year}{2018}), \bibinfo{pages}{66:1--66:36}.
\newblock


\bibitem[\protect\citeauthoryear{Ren, Chen, Li, Ren, Ma, and de~Rijke}{Ren
  et~al\mbox{.}}{2019}]%
        {RenCLR0R19}
\bibfield{author}{\bibinfo{person}{Pengjie Ren}, \bibinfo{person}{Zhumin Chen},
  \bibinfo{person}{Jing Li}, \bibinfo{person}{Zhaochun Ren},
  \bibinfo{person}{Jun Ma}, {and} \bibinfo{person}{Maarten de Rijke}.}
  \bibinfo{year}{2019}\natexlab{}.
\newblock \showarticletitle{RepeatNet: {A} Repeat Aware Neural Recommendation
  Machine for Session-Based Recommendation}. In
  \bibinfo{booktitle}{\emph{AAAI}}. \bibinfo{pages}{4806--4813}.
\newblock


\bibitem[\protect\citeauthoryear{Rendle, Freudenthaler, and
  Schmidt{-}Thieme}{Rendle et~al\mbox{.}}{2010}]%
        {RendleFS10}
\bibfield{author}{\bibinfo{person}{Steffen Rendle}, \bibinfo{person}{Christoph
  Freudenthaler}, {and} \bibinfo{person}{Lars Schmidt{-}Thieme}.}
  \bibinfo{year}{2010}\natexlab{}.
\newblock \showarticletitle{Factorizing personalized Markov chains for
  next-basket recommendation}. In \bibinfo{booktitle}{\emph{WWW}}.
  \bibinfo{pages}{811--820}.
\newblock


\bibitem[\protect\citeauthoryear{Ricci, Rokach, and Shapira}{Ricci
  et~al\mbox{.}}{2015}]%
        {RicciRS15}
\bibfield{editor}{\bibinfo{person}{Francesco Ricci}, \bibinfo{person}{Lior
  Rokach}, {and} \bibinfo{person}{Bracha Shapira}} (Eds.).
  \bibinfo{year}{2015}\natexlab{}.
\newblock \bibinfo{booktitle}{\emph{Recommender Systems Handbook}}.
\newblock \bibinfo{publisher}{Springer}.
\newblock
\showISBNx{978-0-387-85820-3}


\bibitem[\protect\citeauthoryear{Sedhain, Bui, Kawale, Vlassis, Kveton, Menon,
  Bui, and Sanner}{Sedhain et~al\mbox{.}}{2016a}]%
        {SedhainBKVKMBS16}
\bibfield{author}{\bibinfo{person}{Suvash Sedhain}, \bibinfo{person}{Hung~Hai
  Bui}, \bibinfo{person}{Jaya Kawale}, \bibinfo{person}{Nikos Vlassis},
  \bibinfo{person}{Branislav Kveton}, \bibinfo{person}{Aditya~Krishna Menon},
  \bibinfo{person}{Trung Bui}, {and} \bibinfo{person}{Scott Sanner}.}
  \bibinfo{year}{2016}\natexlab{a}.
\newblock \showarticletitle{Practical Linear Models for Large-Scale One-Class
  Collaborative Filtering}. In \bibinfo{booktitle}{\emph{IJCAI}}.
  \bibinfo{pages}{3854--3860}.
\newblock


\bibitem[\protect\citeauthoryear{Sedhain, Menon, Sanner, and Braziunas}{Sedhain
  et~al\mbox{.}}{2016b}]%
        {SedhainMSB16}
\bibfield{author}{\bibinfo{person}{Suvash Sedhain},
  \bibinfo{person}{Aditya~Krishna Menon}, \bibinfo{person}{Scott Sanner}, {and}
  \bibinfo{person}{Darius Braziunas}.} \bibinfo{year}{2016}\natexlab{b}.
\newblock \showarticletitle{On the Effectiveness of Linear Models for One-Class
  Collaborative Filtering}. In \bibinfo{booktitle}{\emph{AAAI}}.
  \bibinfo{pages}{229--235}.
\newblock


\bibitem[\protect\citeauthoryear{Sedhain, Menon, Sanner, and Xie}{Sedhain
  et~al\mbox{.}}{2015}]%
        {SedhainMSX15}
\bibfield{author}{\bibinfo{person}{Suvash Sedhain},
  \bibinfo{person}{Aditya~Krishna Menon}, \bibinfo{person}{Scott Sanner}, {and}
  \bibinfo{person}{Lexing Xie}.} \bibinfo{year}{2015}\natexlab{}.
\newblock \showarticletitle{AutoRec: Autoencoders Meet Collaborative
  Filtering}. In \bibinfo{booktitle}{\emph{WWW}}. \bibinfo{pages}{111--112}.
\newblock


\bibitem[\protect\citeauthoryear{Steck}{Steck}{2019a}]%
        {Steck19c}
\bibfield{author}{\bibinfo{person}{Harald Steck}.}
  \bibinfo{year}{2019}\natexlab{a}.
\newblock \showarticletitle{Collaborative Filtering via High-Dimensional
  Regression}.
\newblock \bibinfo{journal}{\emph{CoRR}}  \bibinfo{volume}{abs/1904.13033}
  (\bibinfo{year}{2019}).
\newblock


\bibitem[\protect\citeauthoryear{Steck}{Steck}{2019b}]%
        {Steck19b}
\bibfield{author}{\bibinfo{person}{Harald Steck}.}
  \bibinfo{year}{2019}\natexlab{b}.
\newblock \showarticletitle{Embarrassingly Shallow Autoencoders for Sparse
  Data}. In \bibinfo{booktitle}{\emph{WWW}}. \bibinfo{pages}{3251--3257}.
\newblock


\bibitem[\protect\citeauthoryear{Steck}{Steck}{2019c}]%
        {Steck19a}
\bibfield{author}{\bibinfo{person}{Harald Steck}.}
  \bibinfo{year}{2019}\natexlab{c}.
\newblock \showarticletitle{Markov Random Fields for Collaborative Filtering}.
  In \bibinfo{booktitle}{\emph{NeurIPS}}. \bibinfo{pages}{5474--5485}.
\newblock


\bibitem[\protect\citeauthoryear{Steck, Dimakopoulou, Riabov, and Jebara}{Steck
  et~al\mbox{.}}{2020}]%
        {SteckDRJ20}
\bibfield{author}{\bibinfo{person}{Harald Steck}, \bibinfo{person}{Maria
  Dimakopoulou}, \bibinfo{person}{Nickolai Riabov}, {and} \bibinfo{person}{Tony
  Jebara}.} \bibinfo{year}{2020}\natexlab{}.
\newblock \showarticletitle{{ADMM} {SLIM:} Sparse Recommendations for Many
  Users}. In \bibinfo{booktitle}{\emph{WSDM}}. \bibinfo{pages}{555--563}.
\newblock


\bibitem[\protect\citeauthoryear{Wang, Ren, Mei, Chen, Ma, and de~Rijke}{Wang
  et~al\mbox{.}}{2019b}]%
        {WangRMCMR19}
\bibfield{author}{\bibinfo{person}{Meirui Wang}, \bibinfo{person}{Pengjie Ren},
  \bibinfo{person}{Lei Mei}, \bibinfo{person}{Zhumin Chen},
  \bibinfo{person}{Jun Ma}, {and} \bibinfo{person}{Maarten de Rijke}.}
  \bibinfo{year}{2019}\natexlab{b}.
\newblock \showarticletitle{A Collaborative Session-based Recommendation
  Approach with Parallel Memory Modules}. In \bibinfo{booktitle}{\emph{SIGIR}}.
  \bibinfo{pages}{345--354}.
\newblock


\bibitem[\protect\citeauthoryear{Wang, Cao, and Wang}{Wang
  et~al\mbox{.}}{2019a}]%
        {WangCW19}
\bibfield{author}{\bibinfo{person}{Shoujin Wang}, \bibinfo{person}{Longbing
  Cao}, {and} \bibinfo{person}{Yan Wang}.} \bibinfo{year}{2019}\natexlab{a}.
\newblock \showarticletitle{A Survey on Session-based Recommender Systems}.
\newblock \bibinfo{journal}{\emph{CoRR}} (\bibinfo{year}{2019}).
\newblock


\bibitem[\protect\citeauthoryear{Wu, Tang, Zhu, Wang, Xie, and Tan}{Wu
  et~al\mbox{.}}{2019}]%
        {WuT0WXT19}
\bibfield{author}{\bibinfo{person}{Shu Wu}, \bibinfo{person}{Yuyuan Tang},
  \bibinfo{person}{Yanqiao Zhu}, \bibinfo{person}{Liang Wang},
  \bibinfo{person}{Xing Xie}, {and} \bibinfo{person}{Tieniu Tan}.}
  \bibinfo{year}{2019}\natexlab{}.
\newblock \showarticletitle{Session-Based Recommendation with Graph Neural
  Networks}. In \bibinfo{booktitle}{\emph{AAAI}}. \bibinfo{pages}{346--353}.
\newblock


\bibitem[\protect\citeauthoryear{Xu, Zhao, Liu, Sheng, Xu, Zhuang, Fang, and
  Zhou}{Xu et~al\mbox{.}}{2019}]%
        {XuZLSXZFZ19}
\bibfield{author}{\bibinfo{person}{Chengfeng Xu}, \bibinfo{person}{Pengpeng
  Zhao}, \bibinfo{person}{Yanchi Liu}, \bibinfo{person}{Victor~S. Sheng},
  \bibinfo{person}{Jiajie Xu}, \bibinfo{person}{Fuzhen Zhuang},
  \bibinfo{person}{Junhua Fang}, {and} \bibinfo{person}{Xiaofang Zhou}.}
  \bibinfo{year}{2019}\natexlab{}.
\newblock \showarticletitle{Graph Contextualized Self-Attention Network for
  Session-based Recommendation}. In \bibinfo{booktitle}{\emph{IJCAI}}.
  \bibinfo{pages}{3940--3946}.
\newblock


\end{thebibliography}

\appendix

\section{Proof of Closed-form Solutions}\label{sec:app_solution}

We provide the detailed derivation of the closed-form solutions in Section~\ref{sec:session_aware_models}.

\subsection{Session-aware Linear Item Similarity Model (SLIS)}

\noindent
To solve the constrained optimization problem, we define a new objective function $\mathcal{L}(\mathbf{B}, \mathbf{\mu})$ by applying a Lagrangian multiplier and a KKT condition:
\begin{equation}
  \label{eq:slis_loss2}
  \begin{aligned}
    \mathcal{L}(\mathbf{B}) = \ & \|\mathbf{W}_\text{full} \odot ( \mathbf{X} - \mathbf{X} \cdot \mathbf{B} )\|_F^2 + \lambda \|\mathbf{B}\|_F^2 \\ & \ \ \text{s.t.} \ \ \texttt{diag}(\mathbf{B}) \le \xi \\
  \end{aligned}
\end{equation}

\begin{equation}
  \label{eq:SLIS_kkt_loss}
  \begin{aligned}
    \mathcal{L}(\mathbf{B}, \mu) = \ & \|\mathbf{W}_\text{full} \odot ( \mathbf{X} - \mathbf{X} \cdot \mathbf{B} )\|_F^2 + \lambda \|\mathbf{B}\|_F^2 \\
    & + 2\mathbf{\mu}^{\top}\texttt{diag}(\mathbf{B} - \xi\mathbf{I}),
  \end{aligned}
\end{equation}
where $\mathbf{\mu} \in \mathbb{R}^{n}$ is the KKT multipliers, which satisfies $\forall \text{i}, \mathbf{\mu}_{i} \geq 0$. For convenience, let $\mathbf{D}_\text{full}$ denote  $\texttt{diagMat}(\mathbf{w}^{2}_\text{full})$. Then, we differentiate $\mathcal{L}(\mathbf{B}, \mathbf{\mu})$ with respect to $\mathbf{B}$ to minimize Eq.~\eqref{eq:SLIS_kkt_loss}:
\begin{align}
    \frac{1}{2}\frac{\partial \mathcal{L} (\mathbf{B},\mu) }{\partial \mathbf{B} }  &  \ = (-\mathbf{X}^{\top}) \mathbf{D}_\text{full} (\mathbf{X} - \mathbf{X} \cdot \mathbf{B}) + \lambda\mathbf{B} + \texttt{diagMat}(\mathbf{\mu}) 
  \label{eq:SLIS_kkt_loss_diff} \\
    & \ =    - \mathbf{X}^{\top} \mathbf{D}_\text{full} \mathbf{X} + \mathbf{X}^{\top} \mathbf{D}_\text{full} \mathbf{X} \cdot \mathbf{B} + \lambda\mathbf{B} + \texttt{diagMat}(\mathbf{\mu})  \nonumber \\
    & \ =   (\mathbf{X}^{\top} \mathbf{D}_\text{full} \mathbf{X} + \lambda\mathbf{I})\mathbf{B} - \mathbf{X}^{\top} \mathbf{D}_\text{full} \mathbf{X} +\texttt{diagMat}(\mathbf{\mu}). \nonumber 
\end{align}

Setting this to 0 and solving by $\mathbf{B}$ gives the optimal $\hat{\mathbf{B}}$ as
\begin{align}
    \hat{\mathbf{B}} & = (\mathbf{X}^{\top} \mathbf{D}_\text{full} \mathbf{X} + \lambda\mathbf{I})^{-1} \cdot  [\mathbf{X}^{\top} \mathbf{D}_\text{full} \mathbf{X}+ \lambda\mathbf{I} - \lambda\mathbf{I} -\texttt{diagMat}(\mathbf{\mu})] \nonumber\\
    & = \hat{\mathbf{P}}[\hat{\mathbf{P}}^{-1} - \lambda\mathbf{I} - \texttt{diagMat}(\mathbf{\mu}) ] \nonumber\\
    & = \mathbf{I} - \hat{\mathbf{P}}[\lambda\mathbf{I} + \texttt{diagMat}(\mathbf{\mu}) ] \nonumber \\ 
    & = \mathbf{I} - \lambda \hat{\mathbf{P}} - \hat{\mathbf{P}}\cdot \texttt{diagMat}(\mathbf{\mu}), 
    \label{eq:SLIS_optimal_B_derivation}
\end{align}
where $\hat{\mathbf{P}} = \left( \mathbf{X}^{\top} \mathbf{D}_\text{full}  \mathbf{X} + \lambda\mathbf{I} \right)^{-1}$.

Also, a KKT multiplier $\mathbf{\mu}_{j}$ is zero only if $\mathbf{B}_{jj} \leq \xi$. Otherwise, $\mathbf{\mu}_{j}$ has a non-zero value. In this case, $\mathbf{\mu}_{i}$ serves to regularize the value of $\mathbf{B}_{jj}$ as $\mathbf{B}_{jj} = \xi$. For $\mathbf{B}_{jj}$, we can develop the the following equation:
\begin{equation}
  \label{eq:regulate_by_mu}
  \mathbf{B}_{jj} \ = \xi = \ 1 -  \lambda \mathbf{P}_{jj} - \mathbf{P}_{jj}\mathbf{\mu}_{j}.
\end{equation}

Finally, $\mathbf{\mu}_{j}$ can be expressed as follows:

\begin{equation}
   \mu_{j} = \frac{(1 - \lambda \mathbf{P}_{jj} - \xi )}{\mathbf{P}_{jj}}
     = \frac{(1 - \xi)}{\mathbf{P}_{jj}} - \lambda.
\end{equation}

Substituting $\mu$ in Eq.~\eqref{eq:SLIS_optimal_B_derivation} and enforcing the non-negative elements in $\hat{\mathbf{B}}$ give $\hat{\mathbf{B}}$:
\begin{equation}
  \label{eq:slis_solution4}
  \hat{\mathbf{B}} = \mathbf{I} -  \hat{\mathbf{P}} \cdot  \texttt{diagMat}(\gamma),
\end{equation}
\begin{equation}
    \label{eq:case_of_gamma_App}
    \gamma_{j} =
    \begin{cases}
        \ \ \ \lambda & \text{if} \ \ \ 1 - \lambda \mathbf{P}_{jj} \le \xi \\
        \ \ \ \frac{1-\xi}{\mathbf{P}_{jj}} & \text{otherwise}. 
    \end{cases}
\end{equation}
Here, $\mathbf{\gamma}$ is a vector defined by $\mathbf{\gamma} = \mathbf{\mu} + \lambda \cdot \mathbf{1}$.

\subsection{Session-aware Linear Item Transition Model (SLIT)}

Given the input matrix $\mathbf{S} \in \mathbb{R}^{m \times n}$ and the output matrix $\mathbf{T} \in \mathbb{R}^{m \times n}$, the objective function is expressed by
\begin{equation}
  \label{eq:slit_loss4}
  \argmin_{\mathbf{B}} \ \mathcal{L}(\mathbf{B}) = \left\|\mathbf{W}_\text{par} \odot ( \mathbf{T} - \mathbf{S} \cdot \mathbf{B} )\right\|_F^2 + \lambda \|\mathbf{B}\|_F^2.
\end{equation}

\noindent
Let $\mathbf{D}_\text{par}$ denote $\texttt{diagMat}(\mathbf{w}^{2}_\text{par})$.
The first-order derivative on Eq.~\eqref{eq:slit_loss4} over $\mathbf{B}$  is then given by
\begin{equation}
\label{eq:B_hat_of_SLIT}
\begin{aligned}
   \frac{1}{2} \cdot \frac{\partial \mathcal{L}}{\partial \mathbf{B} }  &  \ =  \ \ (-\mathbf{S}^{\top}){\mathbf{D}_\text{par}(\mathbf{T} - \mathbf{S} \cdot \mathbf{B})} + \lambda \mathbf{B}, \\
    & \ =  \ \  (\mathbf{S}^{\top} \mathbf{D}_\text{par} \mathbf{S} + \lambda\mathbf{I})\cdot\mathbf{B} - \mathbf{S}^{\top} \mathbf{D}_\text{par} \mathbf{T}.
\end{aligned}
\end{equation}

Letting Eq.~\eqref{eq:B_hat_of_SLIT} to $0$ and solving for $\mathbf{B}$ gives the closed-form solution of Eq.~\eqref{eq:slit_loss4}:
\begin{equation}
  \label{eq:slit_solution4}
  \hat{\mathbf{B}} \ = \ \hat{\mathbf{P}}' \cdot [\mathbf{S}^{\top} \mathbf{D}_\text{par} \mathbf{T}],
\end{equation}
where $\hat{{\mathbf{P}}}' = \left( \mathbf{S}^{\top} \mathbf{D}_\text{par} \mathbf{S} + \lambda \mathbf{I} \right)^{-1}$. 

\subsection{Session-aware Linear Item Similarity/Transition Model (SLIST)}

The objective function of SLIST is expressed as follows:
\begin{equation}
\begin{aligned}
  \label{eq:slist_loss}
  \mathcal{L}(\mathbf{B}) = & \; \alpha \left\|\mathbf{W}_\text{full} \odot ( \mathbf{X} - \mathbf{X} \cdot \mathbf{B} )\right\|_F^2 \\
  & + (1-\alpha) \left\|\mathbf{W}_\text{par} \odot ( \mathbf{T} - \mathbf{S} \cdot \mathbf{B} )\right\|_F^2 + \lambda \|\mathbf{B}\|_F^2,
\end{aligned}
\end{equation}

We differentiate $\mathcal{L}(\mathbf{B})$ with $\mathbf{B}$ gives
\begin{equation}
\label{eq:slist_derivative}
\begin{aligned}
   \frac{1}{2} \frac{\partial \mathcal{L}}{\partial \mathbf{B} }  &  \ =  \ \ (-\alpha\mathbf{X}^{\top}) \mathbf{D}_\text{full} (\mathbf{X} - \mathbf{X} \mathbf{B}) \\  & \ \ \ \  + (-(1-\alpha)\mathbf{S}^{\top}){\mathbf{D}_\text{par}(\mathbf{T} - \mathbf{S} \mathbf{B})} + \lambda \mathbf{B}, \\ 
   &  \ =  \ \ (\alpha\mathbf{X}^{\top}\mathbf{D}_\text{full}\mathbf{X} + (1-\alpha)\mathbf{S}^{\top} \mathbf{D}_\text{par} \mathbf{S} + \lambda\mathbf{I})\cdot \mathbf{B} 
   \\ &  \ \ \ \  - \alpha\mathbf{X}^{\top}\mathbf{D}_\text{full}\mathbf{X} -(1-\alpha)\mathbf{S}^{\top}\mathbf{D}_\text{par}\mathbf{T}. \\
\end{aligned}
\end{equation}

The optimal $\hat{\textbf{B}}$ is derived by
\begin{equation}
\begin{aligned}
    \hat{\mathbf{B}} & = \hat{\mathbf{P}} \cdot [\alpha\mathbf{X}^{\top}\mathbf{D}_\text{full}\mathbf{X} + (1-\alpha) \mathbf{S}^{\top}\mathbf{D}_\text{par}\mathbf{T}] ,\\
    & = \hat{\mathbf{P}} \cdot [\hat{\mathbf{P}}^{-1} - (1-\alpha) \mathbf{S}^{\top}\mathbf{D}_\text{par}\mathbf{S} - \lambda\mathbf{I} + (1-\alpha) \mathbf{S}^{\top}\mathbf{D}_\text{par}\mathbf{T} ], \\
    & = \mathbf{I} - \hat{\mathbf{P}}[\lambda\mathbf{I} + (1-\alpha) \mathbf{S}^{\top}\mathbf{D}_\text{par}(\mathbf{S}-\mathbf{T}) ],  \\
    & = \mathbf{I} - \hat{\mathbf{P}}\lambda - (1-\alpha)\hat{\mathbf{P}} \mathbf{S}^{\top}\mathbf{D}_\text{par}(\mathbf{S}-\mathbf{T}). 
    \label{eq:optimal_B_derivation}
\end{aligned}
\end{equation}
where $\hat{\mathbf{P}} = \left( \alpha\mathbf{X}^{\top}\mathbf{D}_\text{full}\mathbf{X} + (1-\alpha)\mathbf{S}^{\top} \mathbf{D}_\text{par} \mathbf{S} + \lambda\mathbf{I} \right)^{-1}$.

\section{Reproducibility}\label{sec:app_Reproducibility}


\begin{table}
\caption{Optimized hyperparameters of SLIST. $\lambda$ is a L2-weight decay, $\alpha$ is a mix-ratio between SLIS and SLIT. $\delta_\text{pos}$ is a weight decay by item position, $\delta_\text{inf}$ is a weight decay by inference item position and $\delta_\text{time}$ is a weight decay by session recency.}
\label{tab:hyperparameter}
\begin{tabular}{c|cc|ccc}
\toprule
 Dataset & $\lambda$ & $\alpha$  & $\delta_\text{pos}$ & $\delta_\text{inf}$ & $\delta_\text{time}$ \\
\midrule
 YC-1/64 & 10    & 0.4     & 1   & 1     &  4    \\
 YC-1/4  & 10    & 0.2     & 1   & 1     &  8    \\
 DIGI1   & 10    & 0.8     & 1   & 2     &  128  \\
\midrule
 YC      & 10    & 0.2     & 1   & 1     &  8    \\
 DIGI5   & 10    & 0.8     & 1   & 2     &  256  \\
 RR      & 10    & 0.2     & 0.25& 4     &  256  \\
 NOWP    & 10    & 0.8     & 1   & 1     &  128  \\
\bottomrule
\end{tabular}
\end{table}

We set aside a subset from the training set as a validation set, such that the validation contains the same number of days as the test set, \eg, the last $N$ days of sessions from the training set. 

We use a grid search method for hyperparameter searching. 
Since the proposed model guarantees the unique answer given the session via the closed-form solution, it always returns the same recommendations for the same setting. Thus, the hyperparameter is evaluated only once.

Table~\ref{tab:hyperparameter} summarizes all the hyperparameters for SLIST. The performances of the baseline models are taken from the papers~\cite{LudewigJ18, LudewigMLJ19b}. For the hyperparameter setting of each model, we refer to \cite{GargGMVS19, GuptaGMVS19} for the setting of STAN and SR-GNN, respectively. For other models, we refer to~\cite{LudewigJ18,LudewigMLJ19b}\footnote{https://rn5l.github.io/session-rec/umuai/} \footnote{https://rn5l.github.io/session-rec/index.html}. We verify that the performance of baseline models is reproduced with an error of 1--2\% or less in our implementation environments.

\section{Additional Experimental Results}\label{sec:app_additionalresults}

Table~\ref{tab:additional_result} reports additional experimental results with different cut-offs, 5 and 10, for all datasets. We observe a similar tendency as shown in Table~\ref{tab:overallresult}. No competing models show the best performance on all the datasets. For instance, SR-GNN is the best baseline in YC-1/4 and YC-1/64 datasets, and STAN is the best baseline in the other datasets. In contrast, SLIST consistently shows the best or second-best performance up to 12.86\% in HR@20 and 9.62\% in MRR@20 over the best competing model.

\section{Effect of Hyperparameters}\label{sec:app_hyperparameters}
\begin{figure}
\centering
\begin{tabular}{cc}
\includegraphics[width=0.22\textwidth]{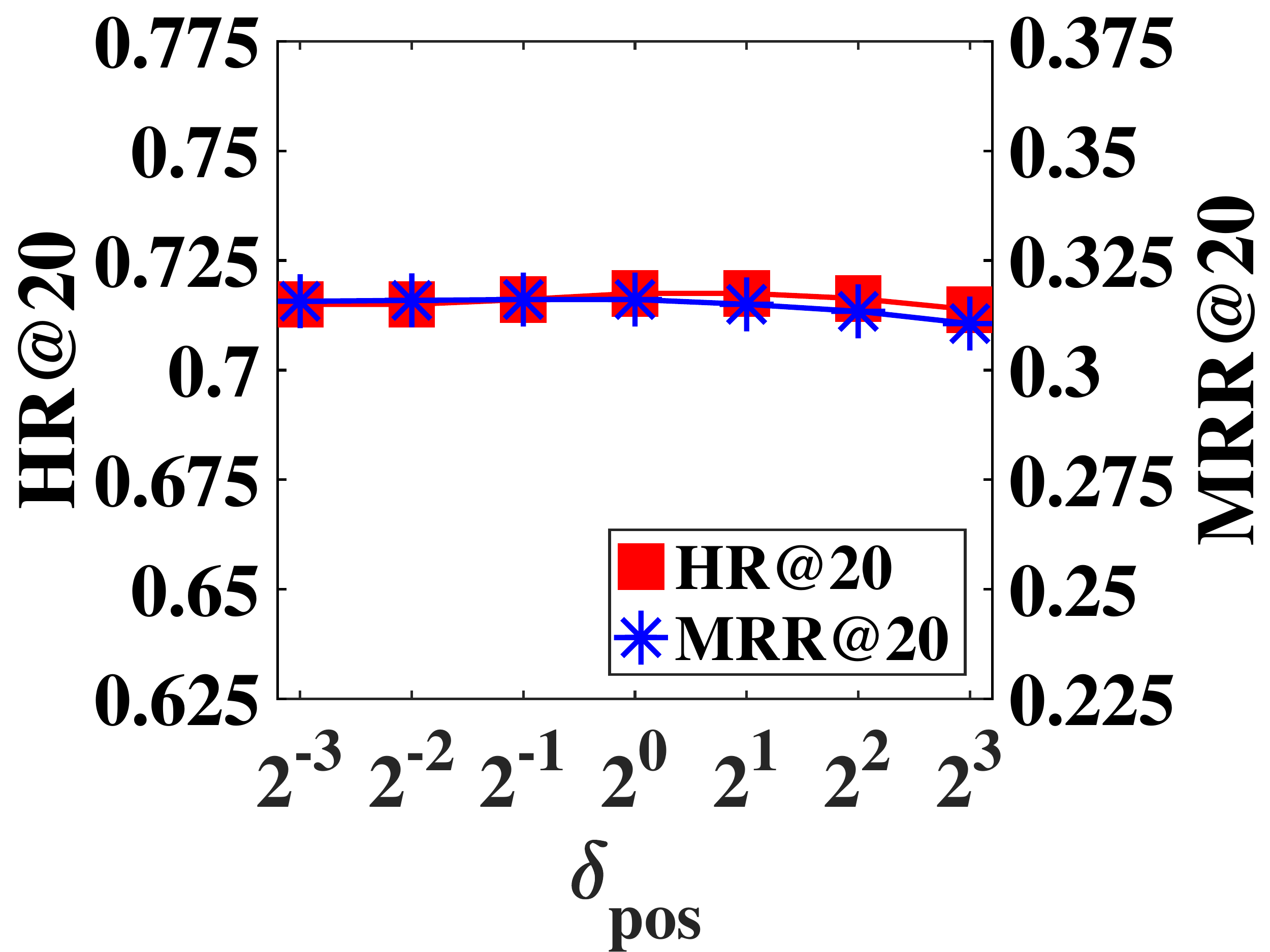} &
\includegraphics[width=0.22\textwidth]{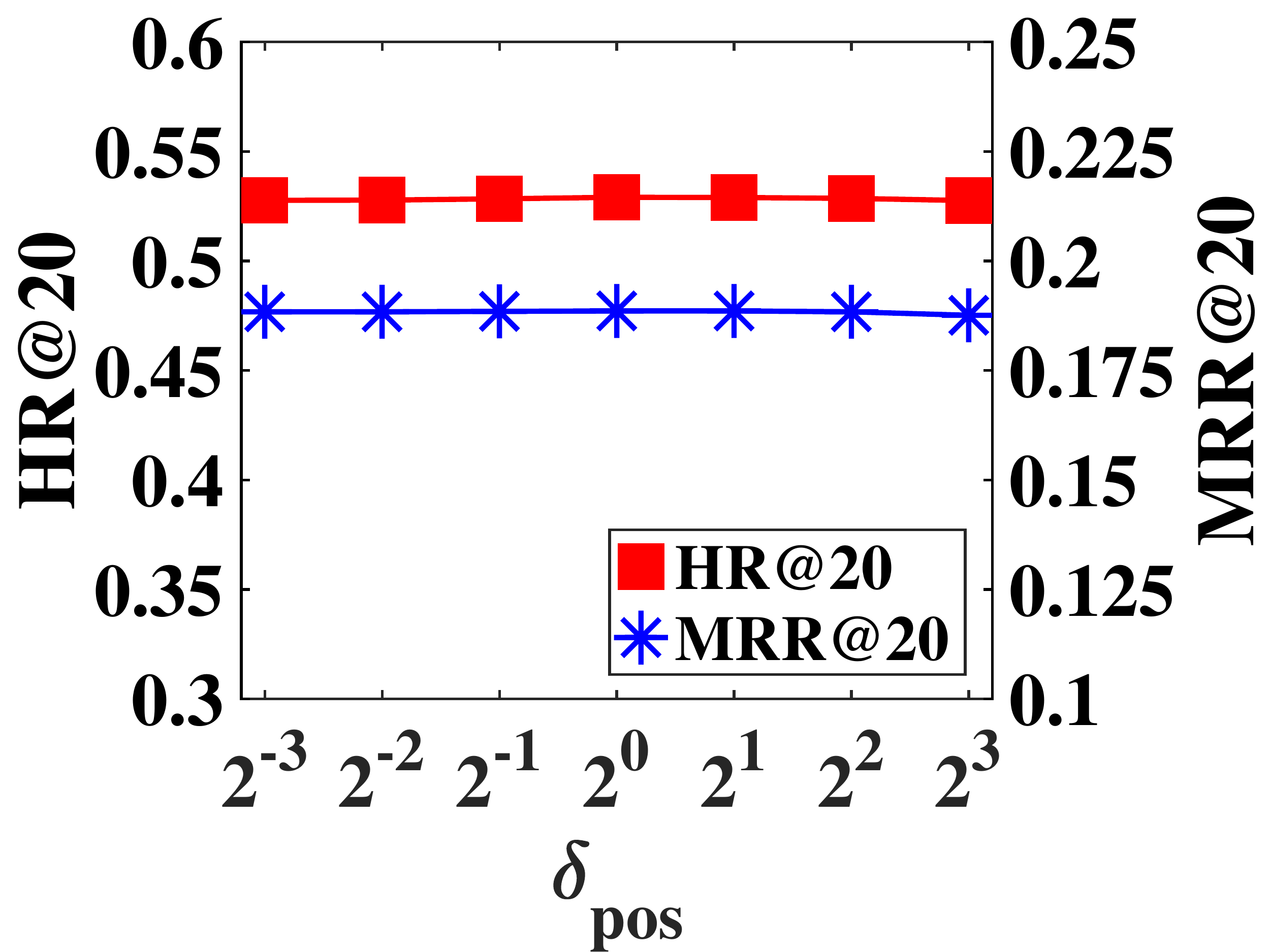} \\
(a) YC-1/4 & (b) DIGI1 \\
\end{tabular}
\vskip -0.1in
\caption{HR@20 and MRR@20 over varying $\delta_\text{pos}$ in SLIST on YC-1/4 and DIGI1.}
\label{fig:train_w}
\end{figure}

\begin{figure}
\centering
\begin{tabular}{cc}
\includegraphics[width=0.22\textwidth]{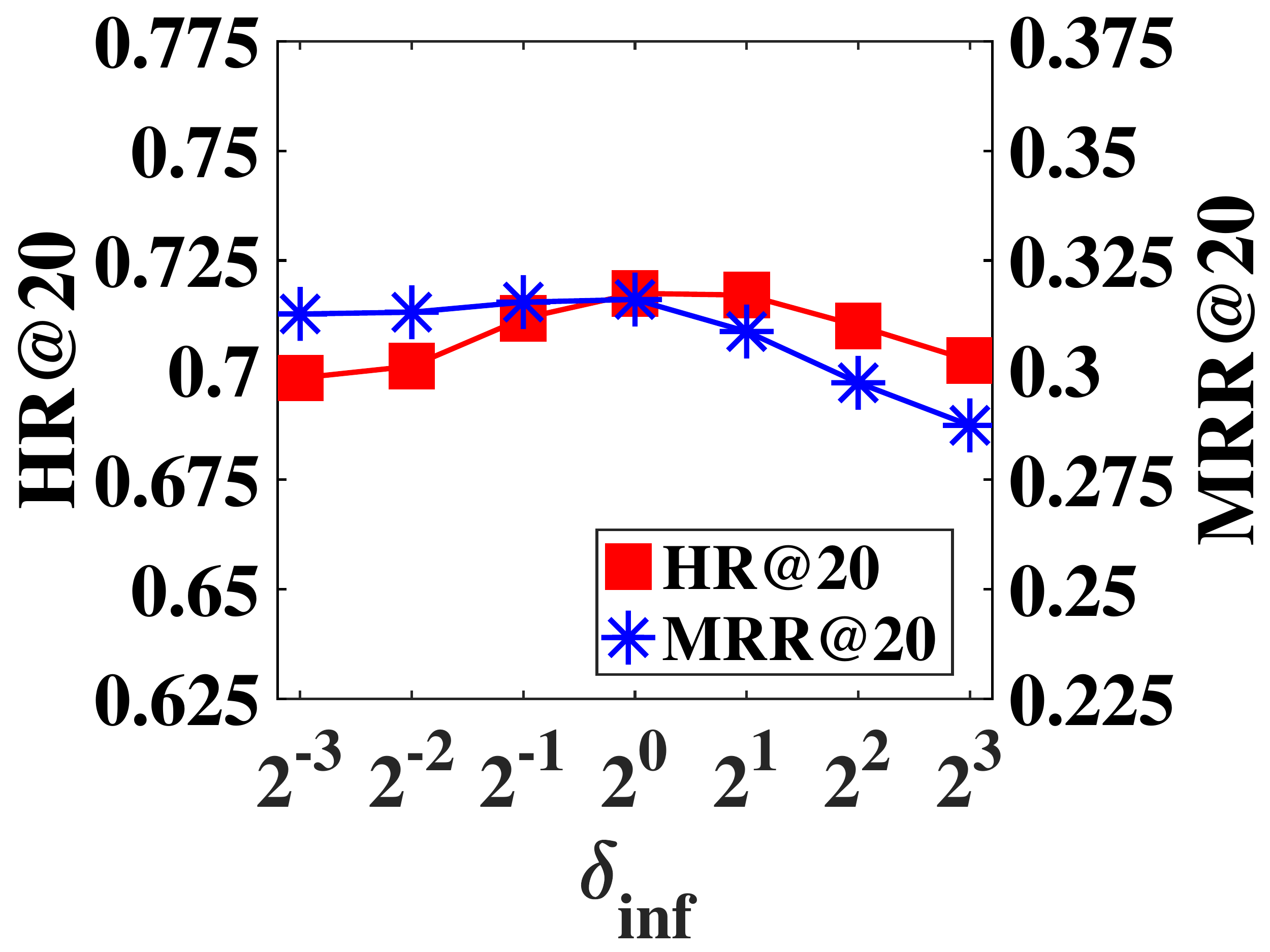} &
\includegraphics[width=0.22\textwidth]{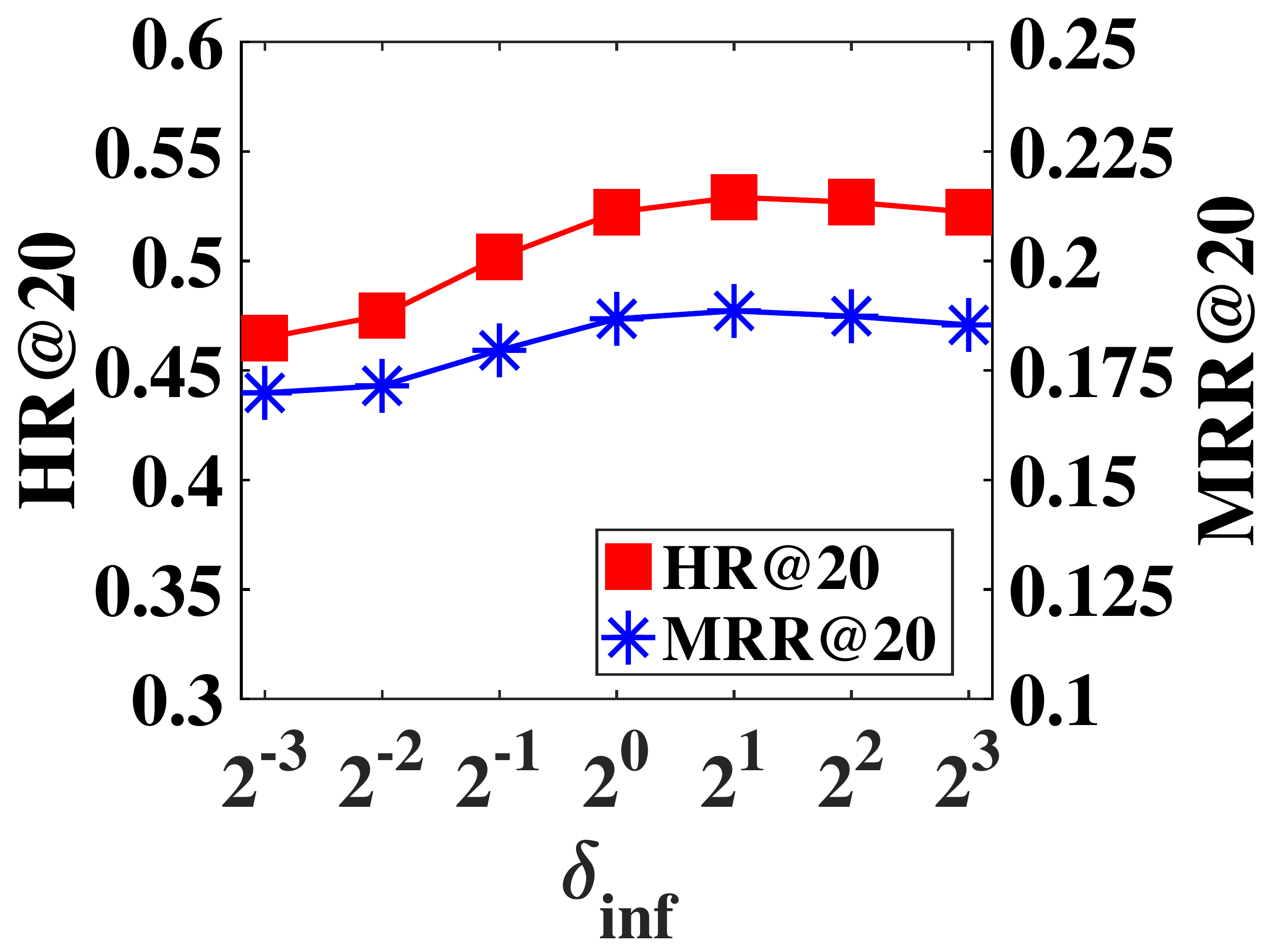} \\
(a) YC-1/4 & (b) DIGI1 \\
\end{tabular}
\vskip -0.1in
\caption{HR@20 and MRR@20 over varying $\delta_\text{inf}$ in SLIST on YC-1/4 and DIGI1.}
\label{fig:predict_w}
\end{figure}

\begin{figure}
\centering
\begin{tabular}{cc}
\includegraphics[width=0.22\textwidth]{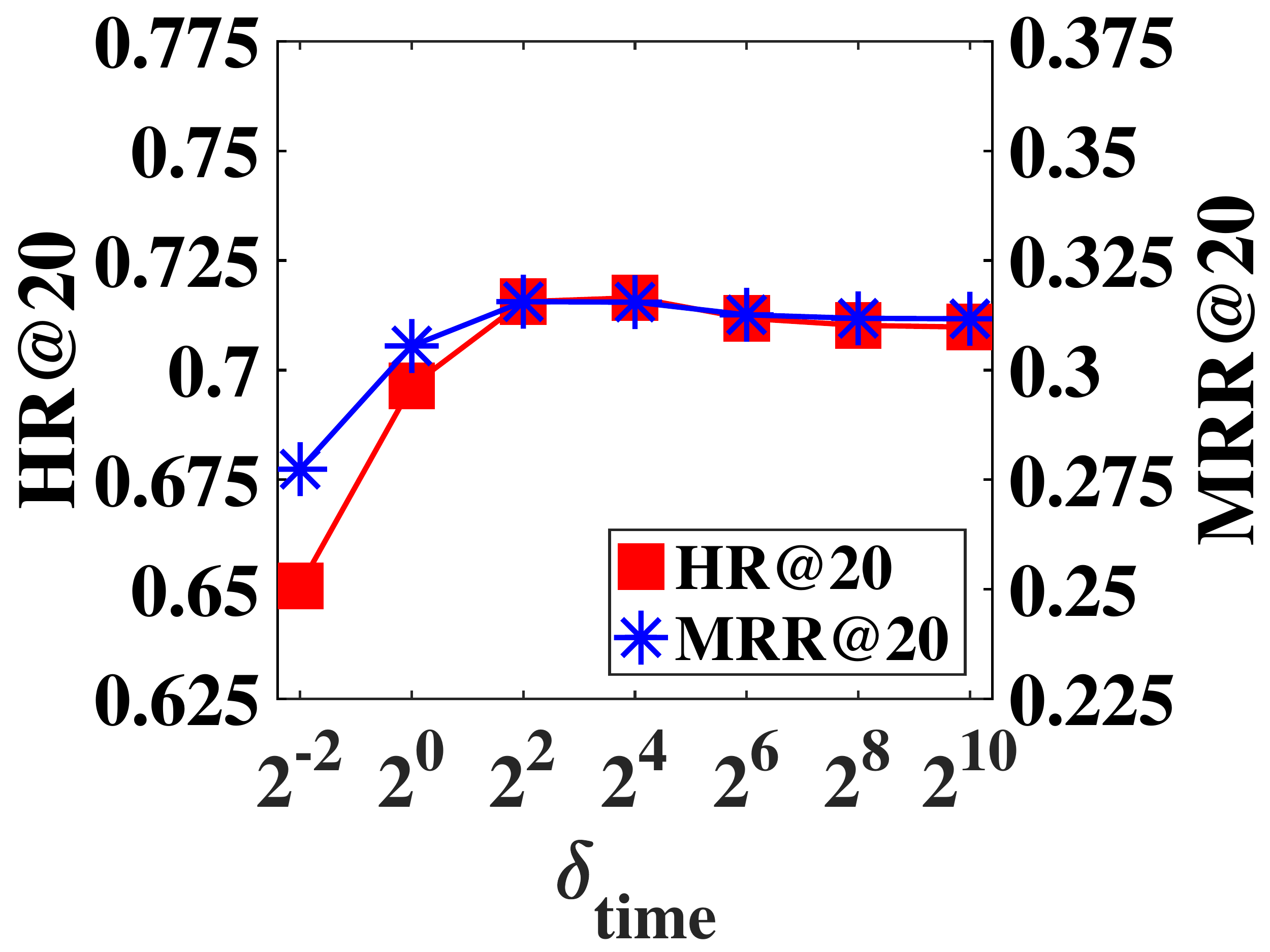} &
\includegraphics[width=0.22\textwidth]{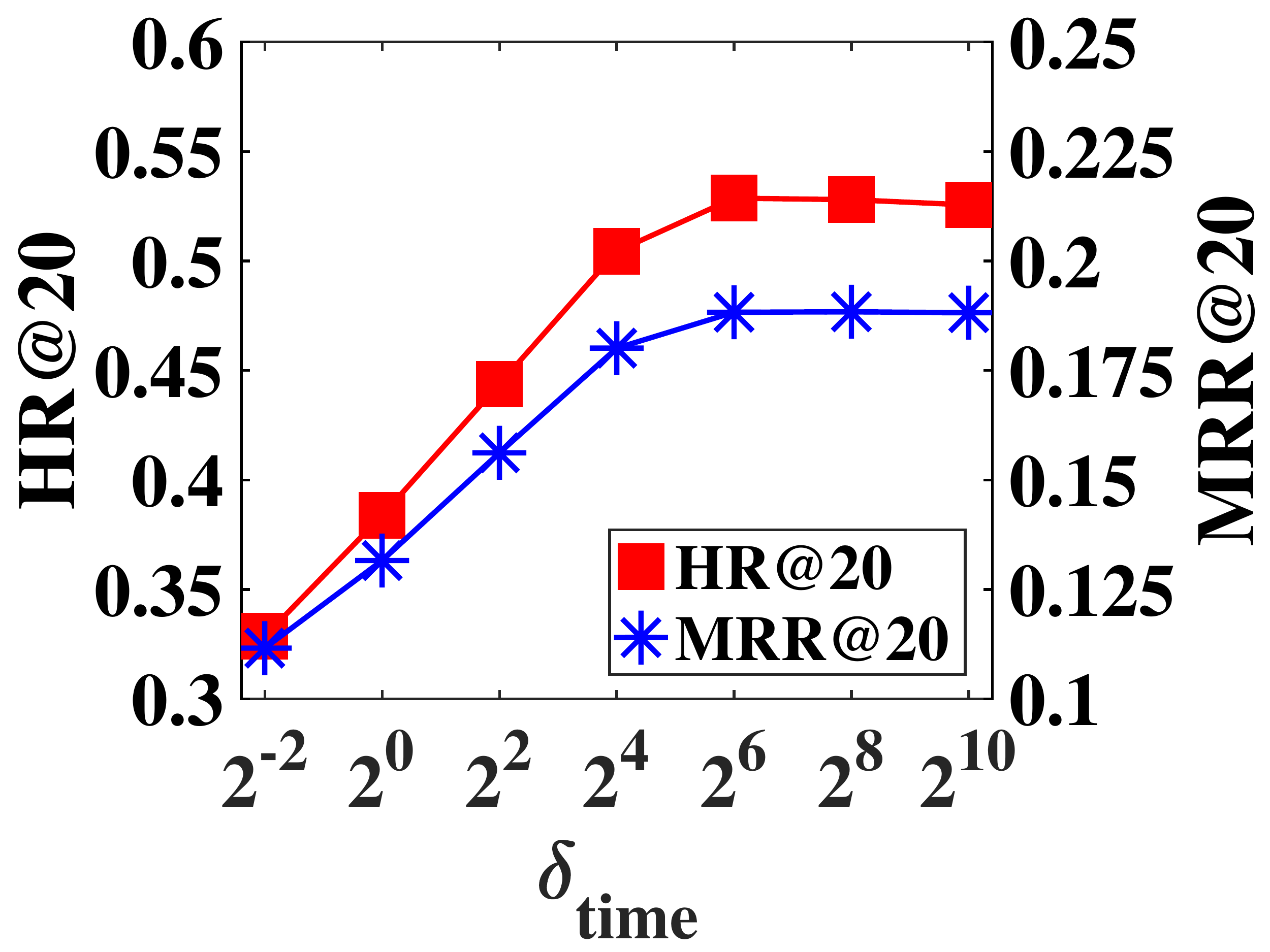} \\
(a) YC-1/4 & (b) DIGI1 \\
\end{tabular}
\vskip -0.1in
\caption{HR@20 and MRR@20 over varying $\delta_\text{time}$ in SLIST on YC-1/4 and DIGI1.}
\label{fig:session_w}
\end{figure}
We conduct additional ablation study in Section~\ref{sec:exp_ablation}. Figures \ref{fig:train_w}, \ref{fig:predict_w}, and \ref{fig:session_w} depict the HR@20 and MRR@20 results on YC-1/4 and DIGI1 datasets over three hyperparameters (\ie, $\delta_\text{pos} , \delta_\text{inf}$, and $\delta_\text{time}$). Figure~\ref{fig:train_w} depicts the effect of the weight decay by item position $\delta_\text{pos}$. It is found that $\delta_\text{pos}$ is less sensitive than the other hyperparameters. Figure~\ref{fig:predict_w} depicts the effect of the weight decay by inference item position $\delta_\text{inf}$. When $\delta_\text{inf} = 2^{0}$ and $\delta_\text{inf} = 2^{1}$, we show the best performance in the YC-1/4 and DIGI1 datasets, respectively. Figure~\ref{fig:session_w} shows the effect of the weight decay by session recency $\delta_\text{time}$. For YC-1/4 and DIGI1 datasets, we show the best performance in $\delta_\text{time} = 2^{2}$ and $\delta_\text{time} = 2^{6}$, respectively.

\begin{table*}[t] \small
\caption{Performance comparison of proposed models and competing models for cut-offs (\ie, 5 and 10), following experimental set-up in \cite{LudewigMLJ19a, LudewigMLJ19b}. Gains indicate how much better the best proposed model is than the best competing model. The best proposed model is marked in \com{\textbf{bold}} and the best baseline model is \fcom{\underline{underlined}}.}
\label{tab:additional_result}
\begin{center}
\begin{tabular}{c|c|cccc|cccc|ccc|c}
\toprule
\multirow{2}{*}{Dataset} & \multirow{2}{*}{Metric} & \multicolumn{4}{c|}{Non-DNN-based models} & \multicolumn{4}{c|}{DNN-based models} & \multicolumn{3}{c|}{Ours} & \multirow{2}{*}{Gains(\%)} \\ 
                         &        & SR       & SKNN     & STAN    & SEASE$^{R}$  & GRU4Rec$^+$   & NARM     & STAMP   & SR-GNN  & SLIS    & SLIT   & SLIST  &                            \\ \hline
\multirow{8}{*}{YC-1/64} & HR@5                    & 0.4606       & 0.3934 & 0.4466       & 0.3007      & 0.4211       & 0.4530       & 0.4536 & \fcom{\ul 0.4701} & 0.4550          & 0.4690          & \com{\textbf{0.4761}} & 1.28                       \\
                         & MRR@5                   & 0.2814       & 0.2263 & 0.2572       & 0.1716      & 0.2510       & 0.2697       & 0.2754 & \fcom{\ul 0.2854} & 0.2574          & \com{\textbf{0.2853}} & 0.2839          & -0.04                      \\
                         & R@5                     & 0.2994       & 0.2834 & 0.3049       & 0.2077      & 0.1954       & 0.3052       & 0.3021 & \fcom{\ul 0.3105} & 0.3092          & 0.3080          & \com{\textbf{0.3149}} & 1.42                       \\
                         & MAP@5                   & 0.0636       & 0.0600 & 0.0644       & 0.0427      & 0.0378       & 0.0645       & 0.0640 & \fcom{\ul 0.0661} & 0.0649          & 0.0657          & \com{\textbf{0.0671}} & 1.51                       \\ \cline{2-14} 
                         & HR@10                   & 0.5751       & 0.5255 & 0.5736       & 0.4212      & 0.5447       & 0.5874       & 0.5793 & \fcom{\ul 0.5976} & 0.5931          & 0.5938          & \com{\textbf{0.6021}} & 0.75                       \\
                         & MRR@10                  & 0.2968       & 0.2440 & 0.2743       & 0.1877      & 0.2676       & 0.2878       & 0.2922 & \fcom{\ul 0.3026} & 0.2760          & \com{\textbf{0.3021}} & 0.3008          & -0.17                      \\
                         & R@10                    & 0.3881       & 0.3809 & 0.4024       & 0.2989      & 0.2988       & 0.4071       & 0.3977 & \fcom{\ul 0.4099} & 0.4110          & 0.4044          & \com{\textbf{0.4123}} & 0.59                       \\
                         & MAP@10                  & 0.0499       & 0.0490 & 0.052        & 0.0382      & 0.0366       & 0.0531       & 0.0514 & \fcom{\ul 0.0536} & 0.0533          & 0.0525          & \com{\textbf{0.0536}} & 0.00                       \\ \hline
\multirow{8}{*}{YC-1/4}  & HR@5                    & 0.4628       & 0.3902 & 0.4455       & 0.3057      & 0.4539       & 0.4627       & 0.4693 & \fcom{\ul 0.4770} & 0.4525          & 0.4789          & \com{\textbf{0.4801}} & 0.64                       \\
                         & MRR@5                   & 0.2818       & 0.2247 & 0.2578       & 0.1732      & 0.2688       & 0.2739       & 0.2819 & \fcom{\ul 0.2894} & 0.2554          & 0.2905          & \com{\textbf{0.2909}} & 0.51                       \\
                         & R@5                     & 0.3031       & 0.2831 & 0.3045       & 0.2111      & 0.2954       & 0.3108       & 0.3082 & \fcom{\ul 0.3134} & 0.3079          & 0.3148          & \com{\textbf{0.3168}} & 1.07                       \\
                         & MAP@5                   & 0.0643       & 0.0599 & 0.0647       & 0.0432      & 0.0617       & 0.0660       & 0.0651 & \fcom{\ul 0.0663} & 0.0642          & 0.0670          & \com{\textbf{0.0675}} & 1.86                       \\ \cline{2-14} 
                         & HR@10                   & 0.5855       & 0.5146 & 0.5757       & 0.4295      & 0.5854       & 0.5955       & 0.5996 & \fcom{\ul 0.6060} & 0.5969          & 0.6086          & \com{\textbf{0.6105}} & 0.75                       \\
                         & MRR@10                  & 0.2983       & 0.2414 & 0.2753       & 0.1897      & 0.2865       & 0.2917       & 0.2994 & \fcom{\ul 0.3068} & 0.2749          & 0.3080          & \com{\textbf{0.3085}} & 0.55                       \\
                         & R@10                    & 0.3961       & 0.3771 & 0.4036       & 0.3032      & 0.3967       & 0.4117       & 0.4084 & \fcom{\ul 0.4134} & 0.4138          & 0.4149          & \com{\textbf{0.4173}} & 0.94                       \\
                         & MAP@10                  & 0.0513       & 0.0485 & 0.0519       & 0.0387      & 0.0518       & 0.0540       & 0.0533 & \fcom{\ul 0.0540} & 0.0537          & 0.0540          & \com{\textbf{0.0543}} & 0.49                       \\ \hline
\multirow{8}{*}{DIGI1}   & HR@5                    & 0.2225       & 0.2699 & \fcom{\ul 0.2867} & 0.169       & 0.2631       & 0.2455       & 0.2306 & 0.2519       & 0.2905          & 0.2499          & \com{\textbf{0.2950}} & 2.90                       \\
                         & MRR@5                   & 0.1244       & 0.1519 & \fcom{\ul 0.1626} & 0.0885      & 0.1507       & 0.1338       & 0.1265 & 0.1420       & 0.1641          & 0.1376          & \com{\textbf{0.1651}} & 1.54                       \\
                         & R@5                     & 0.1673       & 0.2058 & \fcom{\ul 0.2162} & 0.1268      & 0.1981       & 0.1859       & 0.1765 & 0.1887       & 0.2192          & 0.1861          & \com{\textbf{0.2207}} & 2.08                       \\
                         & MAP@5                   & 0.033        & 0.0419 & \fcom{\ul 0.0443} & 0.024       & 0.0393       & 0.0363       & 0.0346 & 0.0373       & 0.0449          & 0.0365          & \com{\textbf{0.0450}} & 1.58                       \\ \cline{2-14} 
                         & HR@10                   & 0.3128       & 0.3767 & \fcom{\ul 0.3942} & 0.2668      & 0.3758       & 0.3619       & 0.3402 & 0.3622       & 0.4042          & 0.3568          & \com{\textbf{0.4078}} & 3.45                       \\
                         & MRR@10                  & 0.1364       & 0.1661 & \fcom{\ul 0.1769} & 0.1014      & 0.1657       & 0.1492       & 0.1410 & 0.1566       & 0.1793          & 0.1518          & \com{\textbf{0.1801}} & 1.81                       \\
                         & R@10                    & 0.2386       & 0.2889 & \fcom{\ul 0.2998} & 0.2049      & 0.2864       & 0.2775       & 0.2609 & 0.2757       & 0.3075          & 0.2692          & \com{\textbf{0.3088}} & 3.00                       \\
                         & MAP@10                  & 0.0281       & 0.0351 & \fcom{\ul 0.0364} & 0.0239      & 0.0342       & 0.0330       & 0.0309 & 0.0328       & 0.0375          & 0.0320          & \com{\textbf{0.0376}} & 3.30                       \\ \midrule
\multirow{8}{*}{YC}      & HR@5                    & 0.4534       & 0.4039 & \fcom{\ul 0.4708} & 0.2893      & 0.4486       & 0.4640       & 0.4593 & 0.4631       & 0.4493          & 0.4679          & \com{\textbf{0.4706}} & -0.04                      \\
                         & MRR@5                   & 0.2799       & 0.2413 & 0.2722       & 0.1628      & 0.2679       & 0.2822       & 0.2812 & \fcom{\ul 0.2851} & 0.2603          & 0.2848          & \com{\textbf{0.2864}} & 0.46                       \\
                         & R@5                     & 0.3234       & 0.3088 & \fcom{\ul 0.338}  & 0.2107      & 0.3208       & 0.3363       & 0.3271 & 0.3275       & 0.3259          & 0.3307          & \com{\textbf{0.3333}} & -1.39                      \\
                         & MAP@5                   & 0.0680       & 0.0646 & \fcom{\ul 0.071}  & 0.0423      & 0.0668       & 0.0706       & 0.0683 & 0.0679       & 0.0669          & 0.0688          & \com{\textbf{0.0694}} & -2.25                      \\ \cline{2-14} 
                         & HR@10                   & 0.5654       & 0.5119 & 0.583        & 0.4025      & 0.5614       & 0.5820       & 0.5753 & \fcom{\ul 0.5860} & 0.5840          & 0.5914          & \com{\textbf{0.5947}} & 1.49                       \\
                         & MRR@10                  & 0.2950       & 0.2559 & 0.2874       & 0.1779      & 0.2832       & 0.2981       & 0.2969 & \fcom{\ul 0.3017} & 0.2787          & 0.3015          & \com{\textbf{0.3032}} & 0.51                       \\
                         & R@10                    & 0.4106       & 0.3932 & 0.4259       & 0.2983      & 0.4087       & \fcom{\ul 0.4298} & 0.4187 & 0.4249       & 0.4263          & 0.4279          & \com{\textbf{0.4305}} & 0.16                       \\
                         & MAP@10                  & 0.0507       & 0.0484 & 0.0529       & 0.0367      & 0.0508       & \fcom{\ul 0.0536} & 0.0519 & 0.0529       & 0.0530          & 0.0531          & \com{\textbf{0.0535}} & -0.19                      \\ \hline
\multirow{8}{*}{DIGI5}   & HR@5                    & 0.1896       & 0.2628 & \fcom{\ul 0.2767} & 0.1555      & 0.2403       & 0.2130       & 0.2039 & 0.2180       & 0.2822          & 0.2178          & \com{\textbf{0.2827}} & 2.17                       \\
                         & MRR@5                   & 0.1073       & 0.1502 & \fcom{\ul 0.1623} & 0.0812      & 0.1426       & 0.1189       & 0.1127 & 0.1241       & \com{\textbf{0.1634}} & 0.1198          & 0.1608          & 0.68                       \\
                         & R@5                     & 0.1407       & 0.2010 & \fcom{\ul 0.2091} & 0.1165      & 0.1830       & 0.1607       & 0.1537 & 0.1643       & \com{\textbf{0.2143}} & 0.1628          & 0.2139          & 2.49                       \\
                         & MAP@5                   & 0.0275       & 0.0407 & \fcom{\ul 0.0424} & 0.0219      & 0.0362       & 0.0312       & 0.0297 & 0.0323       & \com{\textbf{0.0440}} & 0.0316          & 0.0434          & 3.77                       \\ \cline{2-14} 
                         & HR@10                   & 0.2616       & 0.3658 & \fcom{\ul 0.3758} & 0.2453      & 0.3415       & 0.3096       & 0.2924 & 0.3103       & 0.3864          & 0.3161          & \com{\textbf{0.3891}} & 3.54                       \\
                         & MRR@10                  & 0.1169       & 0.1639 & \fcom{\ul 0.1755} & 0.0930      & 0.1559       & 0.1316       & 0.1245 & 0.1363       & \com{\textbf{0.1772}} & 0.1328          & 0.1750          & 0.97                       \\
                         & R@10                    & 0.1970       & 0.2817 & \fcom{\ul 0.2863} & 0.1880      & 0.2613       & 0.2364       & 0.2232 & 0.2368       & 0.2966          & 0.2411          & \com{\textbf{0.2979}} & 4.05                       \\
                         & MAP@10                  & 0.0228       & 0.0340 & \fcom{\ul 0.0343} & 0.0216      & 0.0307       & 0.0275       & 0.0258 & 0.0278       & 0.0358          & 0.0281          & \com{\textbf{0.0359}} & 4.66                       \\ \hline
\multirow{8}{*}{RR}      & HR@5                    & 0.3369       & 0.4581 & \fcom{\ul 0.4866} & 0.1511      & 0.4252       & 0.4230       & 0.3438 & 0.4136       & \com{\textbf{0.4783}} & 0.3759          & 0.4782          & -1.71                      \\
                         & MRR@5                   & 0.2363       & 0.3241 & \fcom{\ul 0.3523} & 0.0873      & 0.3091       & 0.3059       & 0.2404 & 0.2931       & \com{\textbf{0.3457}} & 0.2567          & 0.3380          & -1.87                      \\
                         & R@5                     & 0.2713       & 0.3754 & \fcom{\ul 0.3891} & 0.1235      & 0.3427       & 0.3466       & 0.2981 & 0.3379       & \com{\textbf{0.3842}} & 0.3011          & 0.3832          & -1.26                      \\
                         & MAP@5                   & 0.0548       & 0.0776 & \fcom{\ul 0.0817} & 0.0241      & 0.0696       & 0.0709       & 0.0600 & 0.0687       & \com{\textbf{0.0800}} & 0.0610          & 0.0796          & -2.08                      \\ \cline{2-14} 
                         & HR@10                   & 0.3862       & 0.5264 & \fcom{\ul 0.5462} & 0.2081      & 0.4992       & 0.4922       & 0.4060 & 0.4835       & 0.5438          & 0.4422          & \com{\textbf{0.5469}} & 0.13                       \\
                         & MRR@10                  & 0.2431       & 0.3333 & \fcom{\ul 0.3604} & 0.0949      & 0.3190       & 0.3152       & 0.2488 & 0.3024       & \com{\textbf{0.3545}} & 0.2657          & 0.3473          & -1.64                      \\
                         & R@10                    & 0.3103       & 0.4298 & \fcom{\ul 0.4360} & 0.1697      & 0.4016       & 0.4021       & 0.3477 & 0.3944       & 0.4360          & 0.3532          & \com{\textbf{0.4377}} & 0.39                       \\
                         & MAP@10                  & 0.0344       & 0.0489 & \fcom{\ul 0.0496} & 0.0189      & 0.0447       & 0.0449       & 0.0383 & 0.0441       & 0.0497          & 0.0395          & \com{\textbf{0.0499}} & 0.60                       \\ \hline
\multirow{8}{*}{NOWP}    & HR@5                    & 0.1383       & 0.1033 & 0.1394       & 0.0959      & \fcom{\ul 0.1479} & 0.1161       & 0.1164 & 0.1301       & 0.1231          & 0.1539          & \com{\textbf{0.1633}} & 10.41                      \\
                         & MRR@5                   & 0.0989       & 0.0548 & 0.0769       & 0.0515      & \fcom{\ul 0.0998} & 0.0828       & 0.0809 & 0.0930       & 0.0666          & \com{\textbf{0.1094}} & 0.1032          & 9.62                       \\
                         & R@5                     & 0.0749       & 0.0645 & \fcom{\ul 0.0788} & 0.0695      & 0.0684       & 0.0620       & 0.0001 & 0.0637       & 0.0731          & 0.0806          & \com{\textbf{0.0848}} & 7.61                       \\
                         & MAP@5                   & \fcom{\ul 0.0176} & 0.0130 & 0.0166       & 0.0161      & 0.0142       & 0.0136       & 0.0000 & 0.0135       & 0.0145          & 0.0180          & \com{\textbf{0.0182}} & 3.41                       \\ \cline{2-14} 
                         & HR@10                   & 0.1690       & 0.1686 & \fcom{\ul 0.1889} & 0.1465      & 0.1839       & 0.1457       & 0.1504 & 0.1626       & 0.1850          & 0.1895          & \com{\textbf{0.2132}} & 12.86                      \\
                         & MRR@10                  & 0.1030       & 0.0635 & 0.0835       & 0.0582      & \fcom{\ul 0.1046} & 0.0867       & 0.0854 & 0.0973       & 0.0748          & \com{\textbf{0.1142}} & 0.1098          & 9.18                       \\
                         & R@10                    & 0.1053       & 0.1158 & \fcom{\ul 0.1215} & 0.1115      & 0.0987       & 0.0906       & 0.0003 & 0.0911       & 0.1223          & 0.1152          & \com{\textbf{0.1294}} & 6.50                       \\
                         & MAP@10                  & 0.0166       & 0.0173 & \fcom{\ul 0.0186} & 0.0185      & 0.0131       & 0.0131       & 0.0000 & 0.0122       & 0.0182          & 0.0167          & \com{\textbf{0.0193}} & 3.76                       \\
                         
\bottomrule
\end{tabular}
\end{center}
\end{table*}

\end{document}